\font \eightrm=cmr8

\documentclass[10pt]{amsart}
\usepackage{amssymb}
\usepackage{amsmath}
\usepackage{amsthm}
\usepackage{amscd}
\usepackage[all]{xy}               
\usepackage{epsfig,exscale}
\usepackage{color}
\usepackage{axodraw}
\usepackage{colordvi}

 \newcommand{\nc}{\newcommand}

\newtheorem{thm}{Theorem}
\newtheorem{exam}{Example}

\newtheorem{cor}[thm]{Corollary}

\newtheorem{lem}[thm]{Lemma}
\newtheorem{prop}[thm]{Proposition}
\newtheorem{defn}{Definition}
\newtheorem{rmk}[thm]{Remark}

\def\at1{\begin{array}{c} \ta1\ \\ \end{array}}
\def\Mat31{\begin{array}{c} \td31\ \\ \end{array}}
\def\mat41{\begin{array}{c} \tb2\ \\ \end{array}}
\def\mot43{\begin{array}{c} \th43\ \\ \end{array}}

\def\ta1{{\scalebox{0.25}{ 
\begin{picture}(12,12)(38,-38)
\SetWidth{0.5} \SetColor{Black} \Vertex(45,-33){5.66}
\end{picture}}}}

\def\tb2{{\scalebox{0.25}{ 
\begin{picture}(12,42)(38,-38)
\SetWidth{0.5} \SetColor{Black} \Vertex(45,-3){5.66}
\SetWidth{1.0} \Line(45,-3)(45,-33) \SetWidth{0.5}
\Vertex(45,-33){5.66}
\end{picture}}}}

\def\tc3{{\scalebox{0.25}{ 
\begin{picture}(12,72)(38,-38)
\SetWidth{0.5} \SetColor{Black} \Vertex(45,27){5.66}
\SetWidth{1.0} \Line(45,27)(45,-3) \SetWidth{0.5}
\Vertex(45,-33){5.66} \SetWidth{1.0} \Line(45,-3)(45,-33)
\SetWidth{0.5} \Vertex(45,-3){5.66}
\end{picture}}}}

\def\td31{{\scalebox{0.25}{ 
\begin{picture}(42,42)(23,-38)
\SetWidth{0.5} \SetColor{Black} \Vertex(45,-3){5.66}
\Vertex(30,-33){5.66} \Vertex(60,-33){5.66} \SetWidth{1.0}
\Line(45,-3)(30,-33) \Line(60,-33)(45,-3)
\end{picture}}}}

\def\te4{{\scalebox{0.25}{ 
\begin{picture}(12,102)(38,-8)
\SetWidth{0.5} \SetColor{Black} \Vertex(45,57){5.66}
\Vertex(45,-3){5.66} \Vertex(45,27){5.66} \Vertex(45,87){5.66}
\SetWidth{1.0} \Line(45,57)(45,27) \Line(45,-3)(45,27)
\Line(45,57)(45,87)
\end{picture}}}}

\def\tf41{{\scalebox{0.25}{ 
\begin{picture}(42,72)(38,-8)
\SetWidth{0.5} \SetColor{Black} \Vertex(45,27){5.66}
\Vertex(45,-3){5.66} \SetWidth{1.0} \Line(45,27)(45,-3)
\SetWidth{0.5} \Vertex(60,57){5.66} \SetWidth{1.0}
\Line(45,27)(60,57) \SetWidth{0.5} \Vertex(75,27){5.66}
\SetWidth{1.0} \Line(75,27)(60,57)
\end{picture}}}}

\def\tg42{{\scalebox{0.25}{ 
\begin{picture}(42,72)(8,-8)
\SetWidth{0.5} \SetColor{Black} \Vertex(45,27){5.66}
\Vertex(45,-3){5.66} \SetWidth{1.0} \Line(45,27)(45,-3)
\SetWidth{0.5} \Vertex(15,27){5.66} \Vertex(30,57){5.66}
\SetWidth{1.0} \Line(15,27)(30,57) \Line(45,27)(30,57)
\end{picture}}}}

\def\th43{{\scalebox{0.25}{ 
\begin{picture}(42,42)(8,-8)
\SetWidth{0.5} \SetColor{Black} \Vertex(45,-3){5.66}
\Vertex(15,-3){5.66} \Vertex(30,27){5.66} \SetWidth{1.0}
\Line(15,-3)(30,27) \Line(45,-3)(30,27) \Line(30,27)(30,-3)
\SetWidth{0.5} \Vertex(30,-3){5.66}
\end{picture}}}}

\def\thj44{{\scalebox{0.25}{ 
\begin{picture}(42,72)(8,-8)
\SetWidth{0.5} \SetColor{Black} \Vertex(30,57){5.66}
\SetWidth{1.0} \Line(30,57)(30,27) \SetWidth{0.5}
\Vertex(30,27){5.66} \SetWidth{1.0} \Line(45,-3)(30,27)
\SetWidth{0.5} \Vertex(45,-3){5.66} \Vertex(15,-3){5.66}
\SetWidth{1.0} \Line(15,-3)(30,27)
\end{picture}}}}

\def\ti5{{\scalebox{0.25}{ 
\begin{picture}(12,132)(23,-8)
\SetWidth{0.5} \SetColor{Black} \Vertex(30,117){5.66}
\SetWidth{1.0} \Line(30,117)(30,87) \SetWidth{0.5}
\Vertex(30,87){5.66} \Vertex(30,57){5.66} \Vertex(30,27){5.66}
\Vertex(30,-3){5.66} \SetWidth{1.0} \Line(30,-3)(30,27)
\Line(30,27)(30,57) \Line(30,87)(30,57)
\end{picture}}}}

\def\tj51{{\scalebox{0.25}{ 
\begin{picture}(42,102)(53,-38)
\SetWidth{0.5} \SetColor{Black} \Vertex(61,27){4.24}
\SetWidth{1.0} \Line(75,57)(90,27) \Line(60,27)(75,57)
\SetWidth{0.5} \Vertex(90,-3){5.66} \Vertex(60,27){5.66}
\Vertex(75,57){5.66} \Vertex(90,-33){5.66} \SetWidth{1.0}
\Line(90,-33)(90,-3) \Line(90,-3)(90,27) \SetWidth{0.5}
\Vertex(90,27){5.66}
\end{picture}}}}

\def\tk52{{\scalebox{0.25}{ 
\begin{picture}(42,102)(23,-8)
\SetWidth{0.5} \SetColor{Black} \Vertex(60,57){5.66}
\Vertex(45,87){5.66} \SetWidth{1.0} \Line(45,87)(60,57)
\SetWidth{0.5} \Vertex(30,57){5.66} \SetWidth{1.0}
\Line(30,57)(45,87) \SetWidth{0.5} \Vertex(30,-3){5.66}
\SetWidth{1.0} \Line(30,-3)(30,27) \SetWidth{0.5}
\Vertex(30,27){5.66} \SetWidth{1.0} \Line(30,57)(30,27)
\end{picture}}}}

\def\tl53{{\scalebox{0.25}{ 
\begin{picture}(42,102)(8,-8)
\SetWidth{0.5} \SetColor{Black} \Vertex(30,57){5.66}
\Vertex(30,27){5.66} \SetWidth{1.0} \Line(30,57)(30,27)
\SetWidth{0.5} \Vertex(30,87){5.66} \SetWidth{1.0}
\Line(30,27)(45,-3) \SetWidth{0.5} \Vertex(15,-3){5.66}
\SetWidth{1.0} \Line(15,-3)(30,27) \Line(30,57)(30,87)
\SetWidth{0.5} \Vertex(45,-3){5.66}
\end{picture}}}}

\def\tm54{{\scalebox{0.25}{ 
\begin{picture}(42,72)(8,-38)
\SetWidth{0.5} \SetColor{Black} \Vertex(30,-3){5.66}
\SetWidth{1.0} \Line(30,27)(30,-3) \Line(30,-3)(45,-33)
\SetWidth{0.5} \Vertex(15,-33){5.66} \SetWidth{1.0}
\Line(15,-33)(30,-3) \SetWidth{0.5} \Vertex(45,-33){5.66}
\SetWidth{1.0} \Line(30,-33)(30,-3) \SetWidth{0.5}
\Vertex(30,-33){5.66} \Vertex(30,27){5.66}
\end{picture}}}}

\def\tn55{{\scalebox{0.25}{ 
\begin{picture}(42,72)(8,-38)
\SetWidth{0.5} \SetColor{Black} \Vertex(15,-33){5.66}
\Vertex(45,-33){5.66} \Vertex(30,27){5.66} \SetWidth{1.0}
\Line(45,-33)(45,-3) \SetWidth{0.5} \Vertex(45,-3){5.66}
\Vertex(15,-3){5.66} \SetWidth{1.0} \Line(30,27)(45,-3)
\Line(15,-3)(30,27) \Line(15,-3)(15,-33)
\end{picture}}}}

\def\tp56{{\scalebox{0.25}{ 
\begin{picture}(66,111)(0,0)
\SetWidth{0.5} \SetColor{Black} \Vertex(30,66){5.66}
\Vertex(45,36){5.66} \SetWidth{1.0} \Line(30,66)(45,36)
\Line(15,36)(30,66) \SetWidth{0.5} \Vertex(30,6){5.66}
\Vertex(60,6){5.66} \SetWidth{1.0} \Line(60,6)(45,36)
\SetWidth{0.5}
\SetWidth{1.0} \Line(45,36)(30,6) \SetWidth{0.5}
\Vertex(15,36){5.66}
\end{picture}}}}

\def\tq57{{\scalebox{0.25}{ 
\begin{picture}(81,111)(0,0)
\SetWidth{0.5} \SetColor{Black} \Vertex(45,36){5.66}
\Vertex(30,6){5.66} \Vertex(60,6){5.66} \SetWidth{1.0}
\Line(60,6)(45,36) \SetWidth{0.5}
\SetWidth{1.0} \Line(45,36)(30,6) \SetWidth{0.5}
\Vertex(75,36){5.66} \SetWidth{1.0} \Line(45,36)(60,66)
\Line(60,66)(75,36) \SetWidth{0.5} \Vertex(60,66){5.66}
\end{picture}}}}

\def\tr58{{\scalebox{0.25}{ 
\begin{picture}(81,111)(0,0)
\SetWidth{0.5} \SetColor{Black} \Vertex(60,6){5.66}
\Vertex(75,36){5.66} \SetWidth{1.0} \Line(60,66)(75,36)
\SetWidth{0.5} \Vertex(60,66){5.66}
\SetWidth{1.0} \Line(60,36)(60,66) \Line(60,6)(60,36)
\SetWidth{0.5} \Vertex(60,36){5.66} \Vertex(45,36){5.66}
\SetWidth{1.0} \Line(60,66)(45,36)
\end{picture}}}}

\def\ts59{{\scalebox{0.25}{ 
\begin{picture}(81,111)(0,0)
\SetWidth{0.5} \SetColor{Black}
\Vertex(75,36){5.66} \SetWidth{1.0} \Line(60,66)(75,36)
\SetWidth{0.5} \Vertex(60,66){5.66}
\SetWidth{1.0} \Line(60,36)(60,66) \SetWidth{0.5}
\Vertex(60,36){5.66} \Vertex(45,36){5.66} \SetWidth{1.0}
\Line(60,66)(45,36) \Line(75,6)(75,36) \SetWidth{0.5}
\Vertex(75,6){5.66}
\end{picture}}}}

\def\tt591{{\scalebox{0.25}{ 
\begin{picture}(81,111)(0,0)
\SetWidth{0.5} \SetColor{Black}
\Vertex(75,36){5.66} \SetWidth{1.0} \Line(60,66)(75,36)
\SetWidth{0.5} \Vertex(60,66){5.66}
\SetWidth{1.0} \Line(60,36)(60,66) \SetWidth{0.5}
\Vertex(60,36){5.66} \Vertex(45,36){5.66} \SetWidth{1.0}
\Line(60,66)(45,36) \SetWidth{0.5} \Vertex(45,6){5.66}
\SetWidth{1.0} \Line(45,6)(45,36)
\end{picture}}}}

\nc{\mrm}[1]{{\rm #1}}
\nc{\dirlim}{\displaystyle{\lim_{\longrightarrow}}\,}
\nc{\invlim}{\displaystyle{\lim_{\longleftarrow}}\,}
\nc{\vep}{\varepsilon} \nc{\ep}{\epsilon}

\nc{\mchar}{\mrm{Char}} \nc{\Hom}{\mrm{Hom}} \nc{\id}{\mrm{id}}

\nc{\remark}{\noindent{\bf{Remark:}}}
\nc{\remarks}{\noindent{\bf{Remarks:}}}

 \nc{\delete}[1]{}
 \nc{\grad}[1]{^{({#1})}}
 \nc{\fil}[1]{_{#1}}

\nc{\BA}{{\Bbb A}} \nc{\CC}{{\Bbb C}} \nc{\DD}{{\Bbb D}}
\nc{\EE}{{\Bbb E}} \nc{\FF}{{\Bbb F}} \nc{\GG}{{\Bbb G}}
\nc{\HH}{{\Bbb H}} \nc{\LL}{{\Bbb L}} \nc{\NN}{{\Bbb N}}
\nc{\PP}{{\Bbb P}} \nc{\QQ}{{\Bbb Q}} \nc{\RR}{{\Bbb R}}
\nc{\TT}{{\Bbb T}} \nc{\VV}{{\Bbb V}} \nc{\ZZ}{{\Bbb Z}}
\nc{\Cal}[1]{{\mathcal {#1}}}
\nc{\mop}[1]{\mathop{\hbox {\rm #1} }}
\nc{\mopl}[1]{\mathop{\hbox {\rm #1} }\limits}
\nc{\frakg}{{\frak g}}
\nc{\g}[1]{{\frak {#1}}}
\def \restr#1{\mathstrut_{\textstyle |}\raise-8pt\hbox{$\scriptstyle #1$}}
\def \srestr#1{\mathstrut_{\scriptstyle |}\hbox to
  -1.5pt{}\raise-4pt\hbox{$\scriptscriptstyle #1$}}
\nc{\wt}{\widetilde} \nc{\wh}{\widehat}

\nc{\redtext}[1]{\textcolor{red}{#1}}
\nc{\bluetext}[1]{\textcolor{blue}{#1}}

\nc\fleche[1]{\mathop{\hbox to #1 mm{\rightarrowfill}}\limits}
\def\semi{\mathrel{\times}\kern -.85pt\joinrel\mathrel{\raise 1.4pt\hbox{${\scriptscriptstyle |}$}}}

\begin{document}

\title[On matrix differential equations in the Hopf algebra of renormalization]
      {On matrix differential equations in the \\ Hopf algebra of renormalization}

\author{Kurusch Ebrahimi-Fard}
\address{I.H.\'E.S.,
         Le Bois-Marie,
         35, Route de Chartres,
         F-91440 Bures-sur-Yvette, France}
         \email{kurusch@ihes.fr}
         \urladdr{http://www.th.physik.uni-bonn.de/th/People/fard/}

\author{Dominique Manchon}
\address{Universit\'e Blaise Pascal,
         C.N.R.S.-UMR 6620,
         63177 Aubi\`ere, France}
         \email{manchon@math.univ-bpclermont.fr}
         \urladdr{http://math.univ-bpclermont.fr/~manchon/}

\date{June 26, 2006\\ \noindent {\footnotesize{${}\phantom{a}$ 2001 PACS Classification:
03.70.+k, 11.10.Gh, 02.10.Hh}} }

\begin{abstract}
We establish Sakakibara's differential equations \cite{Sa04} in a
matrix setting for the counter term (respectively renormalized
character) in Connes--Kreimer's Birkhoff decomposition in any
connected graded Hopf algebra, thus including Feynman rules in
perturbative renormalization as a key example.
\end{abstract}

\maketitle

\tableofcontents


\section{Introduction}
\label{sect:intro}

Quantum field theory (QFT) unifies the fundamental principles of
special relativity and quantum theory and provides the appropriate
physical framework to describe phenomena at the smallest length
scales respectively highest energies. Its mathematical structure
is far from being as simple as that of its basic constituents.
Moreover, up to now, perturbation theory is the most successful
quantitative and qualitative approach to QFT. Although general
agreement between theoretically predicted results in the
perturbative regime of QFT and those experimentally measured has
reached a satisfactory status, a truly non-perturbative
understanding of the physics of quantum phenomena is mandatory,
both for future advancements in terms of fundamental as well as
calculational problems.

Perturbative QFT consist of two fundamental ingredients, the gauge
principle and the concept of renormalization. The latter consists
of an arbitrary regularization prescription, which parameterizes
ultraviolet divergencies appearing in Feynman amplitudes and
thereby renders them formally finite, together with a specific
subtraction rule of those ill-defined expressions dictated by
physical principles. Whereas both the gauge principle and the
concept of renormalization experienced a rich development in
theoretical physics, the former especially came to the fore of
mathematical research with rich interactions between
mathematicians and physicists. However, the latter suffered from
the lack of an equally strong development of its mathematical
aspects.

Kreimer's recent findings~\cite{KreimerHopf,KreimerChen} mark a
turning point in this context. He discovered a mathematical
structure underlying renormalization in perturbative quantum field
theory in terms of connected graded commutative Hopf algebras.
Feynman rules are interpreted as Hopf algebra characters which
associate to each Feynman graph its corresponding amplitude.

The concept of regularization in general introduces non-physical
parameters. This process changes the nature of Feynman rules
drastically, i.e., from linear multiplicative maps into the
underlying base field, to algebra morphisms with image in a
commutative unital algebra, e.g., Laurent series in dimensional
regularization. Hence we identify regularized Feynman rules with a
particular subclass of such maps from the Hopf algebra of Feynman
graphs into a commutative unital algebra dictated by the
regularization scheme.

Connes and Kreimer extended the results on the Hopf algebraic
approach to perturbative renormalization by establishing the Hopf
algebra of Feynman graphs including the concept of the
renormalization group~\cite{CK1,CK2,CK3,CK4}. Moreover, Connes and
Kreimer formulated in this picture the intricate process of
perturbative renormalization in terms of an algebraic Birkhoff
decomposition of regularized Feynman rules, using the minimal
subtraction scheme in dimensional regularization.

In \cite{EG05,EGGV} it was shown how to organize the combinatorics
of renormalization in terms of (pro-)nil- and unipotent triangular
matrix representations with entries in a commutative Rota--Baxter
algebra. A simple factorization of such matrices was derived using
explicit non-recursive equations containing the renormalization
scheme operator. This simple matrix decomposition offers a
transparent picture of the process of renormalization in terms of
the factorization of Feynman rules matrices.\smallskip

In this work we would like to further develop the matrix calculus
approach to perturbative renormalization in the abstract context
of connected graded Hopf algebras. Any left coideal gives rise to
a representation of the group of characters of the Hopf algebra by
lower triangular unipotent matrices, the size of which being given
by the dimension of the coideal. We investigate the matrix
representation of two fundamental concepts which can be defined in
this purely algebraic framework: the renormalization group and the
beta-function. We retrieve then M.~Sakakibara's differential
equations involving the beta-function, giving to his approach the
firm ground of triangular matrix calculus.\smallskip

Before starting we should point the reader to the following papers
\cite{CaswellKennedy82,Collins06,Delamotte04,EK05,FG05,Kreimer3,Ma01,tHV73}
and books~\cite{Collins84,FGV01,IZ80,KreimerBuch,Muta87,Vasilev04}
which are useful as introductory references, both with respect to
perturbative QFT and renormalization theory, as well as its
recently discovered Hopf-algebraic structures. Also, some readers
may find it stimulating to leaf through the books by
Brown~\cite{Brown93} and Schweber~\cite{Schweber94} as well as the
more recent one by Kaiser~\cite{Kaiser05} for some
scientific-historical perspectives on Feynman graphs in QFT and
renormalization theory. Schwinger's collection of
reprints~\cite{Schwinger58} contains many of the original articles
marking the beginning of modern perturbative QFT and
renormalization theory. Comprehensive treatments of Hopf algebras
can be found in~\cite{Abe80,Sw69}, see also the paper by
Bergman~\cite{Berg85}. Other useful references
are~\cite{ChariPressley95,FGV01,Kassel95,Majid95,ShSt93}. Hopf
algebras in the context of combinatorics appeared in the work of
Rota~\cite{Rota78}, and Joni and Rota~\cite{JoniRota79}, see also
\cite{FG05,NiSw1982,Schmitt95,SD97}.\\

Let us briefly outline the organization of this paper. In section
\ref{sect:SetUp}, after reminding Connes--Kreimer's Birkhoff
decomposition of characters in the most general context of
connected filtered Hopf algebras, we define the matrix
representation associated with a left coideal, along the lines of
\cite{EG05}, and write down the matrix counterpart of the Birkhoff
decomposition. In section \ref{sect:matrix}, we first define the
renormalization group and the beta-function in the context of
connected graded Hopf algebras, along the lines of \cite{CK2} and
\cite{Ma01}, and then we describe the matrix counterparts of these
notions. The key point is that the grading biderivation $Y$ of the
Hopf algebra can be represented by a diagonal matrix. The
semidirect product of the group of characters with the associated
one-parameter group of automorphisms can then be represented by
(non-unipotent) lower-triangular matrices. The two last
subsections are devoted to a careful rewriting of some important
results of M.~Sakakibara (\cite{Sa04}) in the matrix
representation, yielding matrix differential equations for the
beta-function.


\section{The general set up}
\label{sect:SetUp}

In the sequel $k$ denotes the ground field with $char(k)=0$ over
which all algebraic structures are defined. Here the term algebra
always means unital associative $k$-algebra, denoted by the triple
$(\Cal A,m_{\Cal A},\eta_{\Cal A})$, where $\Cal A$ is a
$k$-vector space with a product $m_{\Cal A}: \Cal A \otimes \Cal A
\to \Cal A$ and a unit map $\eta_{\Cal A}: k \to \Cal A$.
Similarly for coalgebras over $k$, denoted by the triple $(\Cal
C,\Delta_{\Cal C},\epsilon_{\Cal C})$, where the coproduct map
$\Delta_{\Cal C}: \Cal C \to \Cal C \otimes \Cal C$ is
coassociative and $\epsilon_{\Cal C}: \Cal C \to k$ denotes the
counit map. A subspace $\Cal J \subset \Cal C$ is called a left
coideal if $\Delta_{\Cal C}(\Cal J) \subset \Cal C \otimes \Cal
J$. A Hopf algebra, denoted by $(\Cal H,m_{\Cal H},\eta_{\Cal
H},\Delta_{\Cal H},\epsilon_{\Cal H},S)$, is a bialgebra together
with the antipode $S: \Cal H \to \Cal H$, that is, it consists of
an algebra and coalgebra structure in a compatible way and $S$ is
a $k$-linear map on $\Cal H$ satisfying the Hopf algebra
axioms~\cite{Abe80,Sw69}. In the following we omit subscripts for
notational transparency if there is no danger of confusion, and
denote algebras, coalgebras and Hopf algebras simply by $\Cal A$,
$\Cal C$ and $\Cal H$, respectively.


\subsection{Connected filtered Hopf algebra}
\label{ssect:cfHopfAlg}

Let $\Cal H$ be a connected filtered bialgebra:
$$
    k=\Cal H^{(0)} \subset \Cal H^{(1)} \subset \cdots \subset \cdots
      \Cal H^{(n)} \subset \cdots, \hskip 6mm \bigcup_{n\ge 0} \Cal H^{(n)}=\Cal H,
$$
and let $\Cal A$ be any commutative algebra. The space $\Cal
L(\Cal H,\Cal A)$ of linear maps from $\Cal H$ to $\Cal A$
together with the convolution product $f \star g := m_{\Cal A}
\circ (f \otimes g) \circ \Delta$, $ f,g \in \Cal L$~:
 \allowdisplaybreaks{
\begin{equation*}
   \mathcal{H} \xrightarrow{\Delta}
   \mathcal{H} \otimes \mathcal{H}
   \xrightarrow{f \otimes g} {\Cal A} \otimes {\Cal A}
   \xrightarrow{m_{\Cal A}} {\Cal A},
\end{equation*}}
is an algebra with unit $e:=\eta_{\Cal A} \circ \epsilon$. For any
$x \in \Cal H^{(n)}$ we have, using a variant of Sweedler's
notation \cite{Sw69}:
\begin{equation}\label{coprod2}
    \Delta(x) = x \otimes 1 + 1 \otimes x + \sum_{(x)} x'\otimes x'',
\end{equation}
where the filtration degrees of $x'$ and $x''$ are strictly
smaller than $n$. Recall that by definition we call an element $x
\in \Cal H$ primitive if:
$$
    \bar{\Delta}(x) := \Delta(x) -  x \otimes 1 - 1 \otimes x = 0.
$$
The convolution product on $\Cal L(\Cal H,\Cal A)$ writes then
with Sweedler's notation:
\begin{equation}\label{convolution2}
    (f \star g)(x) = f(x) g(1) + f(1)g(x) + \sum_{(x)} f(x') g(x'') \in \Cal A.
\end{equation}
The filtration of $\Cal H$ implies a decreasing filtration on
$\Cal L(\Cal H,\Cal A)$ in terms of $\Cal L^n:=\{f \in \Cal L
\big{|}\ f{\restr{\Cal H^{(n-1)}}}=0 \}$ and $\Cal L(\Cal H,\Cal
A)$ is complete with respect to the induced topology~\cite{Ma01}.
The subset $\g g_0:=\Cal L^1 \subset \Cal L(\Cal H,\Cal A)$ of
linear maps $\alpha$ that send the bialgebra unit to zero,
$\alpha(1)=0$, forms a Lie algebra in $\Cal L(\Cal H,\Cal A)$. The
exponential:
$$
    \exp^\star(\alpha) = \sum_k \frac{1}{k!}\alpha^{\star k}
$$
makes sense and is a bijection from $\g g_0$ onto the group
$G_0=e+\g g_0$ of linear maps $\gamma$ that send the bialgebra
unit to the algebra unit, $\alpha(1)=1_{\Cal A}$ \cite{Ma01}.

An infinitesimal character with values in $\Cal A$ is a linear map
$\xi \in \Cal L(\Cal H,\Cal A)$ such that for $x,y \in \Cal H$~:
\begin{equation}\label{infinitesimal}
    \xi(xy) = \xi(x)e(y) + e(x)\xi(y).
\end{equation}
We denote by $\g g_{\Cal A} \subset \g g_0$ the linear space of
infinitesimal characters. We call an $\Cal A$-valued map $\rho$ in
$\Cal L(\Cal H,\Cal A)$ a character if for $x,y \in \Cal H$~:
\begin{equation}\label{character}
    \rho(xy) = \rho(x)\rho(y),
\end{equation}
The set of such unital algebra morphisms is denoted by $G_{\Cal A}
\subset G_0$.

Let us now assume that $\Cal A$ is a commutative algebra. It is
easily verified then, see for instance~\cite{Ma01}, that the set
$G_{\Cal A}$ of characters from $\Cal H$ to $\Cal A$ forms a group
for the convolution product, in fact it is the pro-unipotent
{\sl{group\footnote{It is more precisely a group scheme, i.e. a
functor $\Cal A \mapsto G_{\Cal A}$ from $k$-algebras to groups.}
of $\Cal A$-valued morphisms on the bialgebra $\Cal H$}\/}{}. And
$\g g_{\Cal A}$ in $\g g_0$ is the corresponding pro-nilpotent Lie
algebra. The exponential map $\exp^{\star}$ restricts to a
bijection between $\g g_{\Cal A}$ and $G_{\Cal A}$. The neutral
element $e:=\eta_{\Cal A}\circ \epsilon$  in $G_{\Cal A}$ is given
by $e(1)=1_{\Cal A}$ and $e(x)=0$ for $x \in \mop{Ker}\epsilon$.
The inverse of $\varphi \in G_{\Cal A}$ is given by composition
with the antipode $S$:
\begin{equation}\label{inverse}
    \varphi^{\star -1} = \varphi \circ S.
\end{equation}

Recall that the antipode $S: \Cal H \to \Cal H$ is the inverse of
the identity for the convolution product on $\Cal L(\Cal H,\Cal
H)$~:
\begin{equation} \label{antipode}
    S  \star Id =  m \circ (S \otimes Id) \circ \Delta = \eta \circ \epsilon = Id \star S.
\end{equation}
It always exists in a connected filtered bialgebra, hence any
connected filtered bialgebra is a {\sl{connected filtered Hopf
algebra}\/}. The antipode is defined by:
\begin{equation}
  \label{def:antipode1}
  S = \sum_{n\ge 0}(\eta \circ \epsilon - Id)^{\star n}.
\end{equation}
Recall that $\Delta^{(0)} := Id$ and for $n>0$
$\Delta^{(n)}:=(\Delta^{(n-1)} \otimes Id) \circ \Delta$.
Equations~(\ref{antipode}) imply the following recursive formulas
for the antipode starting with $S(1)=1$ and for $x \in
\mop{Ker}\epsilon$:
 \allowdisplaybreaks{
\begin{eqnarray}
    \label{antipode2}
    S(x) &= -x- \displaystyle\sum_{(x)}S(x')x'',\\
    S(x) &= -x- \displaystyle\sum_{(x)}x'S(x'').
\end{eqnarray}}

\begin{exam}{\rm{{\bf{Toy-model of decorated non-planar rooted trees}}:
As a guiding example we will use the Hopf algebra of non-planar
rooted trees established by Kreimer~\cite{KreimerChen}. It
provides the role model for the Hopf algebraic formulation of
perturbative renormalization~\cite{CK1}. In fact, linear
combinations of --decorated-- non-planar rooted trees naturally
encode the hierarchical structure of divergencies of a Feynman
graph. Each vertex of a rooted tree represents a primitive
divergence indicated by a decoration with a primitive one-particle
(1PI) irreducible Feynman graph. Edges connecting such vertices
encode thereby the nesting of subdivergencies, i.e., proper 1PI
subgraphs sitting inside another 1PI graph. The root vertex is the
overall divergence which contains those subdivergencies. Linear
combinations of such decorated rooted trees may represent Feynman
graphs with overlapping divergence
structures~\cite{KreimerOver}.\smallskip

By definition a rooted tree $t$ is made out of vertices and
nonintersecting oriented edges, such that all but one vertex have
exactly one incoming line. We denote the set of vertices and edges
of a rooted tree by $V(t)$, $E(t)$ respectively. The root is the
only vertex with no incoming line. We draw the root on top of the
tree. Let $T$ denote the set of isomorphic classes of rooted
trees. The empty tree is denoted by $1_{\mathcal{T}}$:
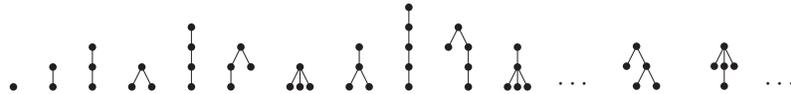
$$
 \ta1           \;\;\;\;
 \tb2           \;\;\;\;
 \tc3           \;\;\;\;
 \td31          \;\;\;\;
 \te4           \;\;\;\;
 \tf41          \;\;\;\;
 \th43          \;\;\;\;
 \thj44          \;\;\;\;
 \ti5           \;\;\;\;
 \tj51         \;\;\;\;
 \tm54         \;\;\; \cdots \;\;\; \tp56  \;\;\; \tr58 \;\;\; \cdots
$$
Let $\mathcal{T}$ be the $k$-vector space generated by $T$, which
is graded by the number of vertices, $\deg(t):=|t|:=|V(t)|$, $t
\in \Cal T$, with the convention that $\deg(1_\mathcal{T})=0$. Let
$\mathcal{H}_\mathcal{T}$ be the graded commutative polynomial
algebra of finite type over $k$ generated by $\mathcal{T}$,
$\mathcal{H}_\mathcal{T}:=k[\mathcal{T}]=\bigoplus_{n\geq 0}
k\mathcal{H}\grad{n}$. Monomials of trees are called forests. We
extend $\deg(t_1 \dots t_n):=\sum_{i=1}^n \deg(t_i)$.

We will define a coalgebra structure on $\mathcal{T}$. The
coproduct is defined in terms of {\it{cuts}} $c \subset E(t)$ on a
tree $t \in \mathcal{T}$. A {\it{primitive cut}} is the removal of
a single edge, $|c|=1$, from the tree $t$. The tree $t$ decomposes
into two parts, denoted by the pruned part $P_c(t)$ and the rooted
part $R_c(t)$, where the latter contains the original root vertex.
An {\it{admissible cut}} of a rooted tree $t$ is a set of
primitive cuts, $|c| \geq 1$, such that along the unique path from
the root to any vertex of $t$ one encounters at most one cut.

Let $C_t$ be the set of all admissible cuts of the rooted tree $t
\in \mathcal{T}$. We exclude the empty cut $c^{(0)}$:
$P_{c^{(0)}}(t)= \emptyset$, $R_{c^{(0)}}(t)=t$, and the full cut
$c^{(1)}$: $P_{c^{(1)}}(t)= t$, $R_{c^{(1)}}(t)=\emptyset$. Also
let $C_t^{(01)}$ be $C_t \cup \{c^{(0)}(t),c^{(1)}(t)\}.$ The
{\it{coproduct}} is defined by:
 \begin{equation}
    \Delta(t) := t \otimes 1_\mathcal{T} + 1_\mathcal{T} \otimes t + \sum_{c \in C_t} P_{c}(t) \otimes R_{c}(t)
               = \sum_{c \in C^{(01)}_t} P_{c}(t) \otimes R_{c}(t).
    \label{coprod1}
 \end{equation}
We shall call the rooted tree $R_{c}(t)$ the cotree (or cograph)
corresponding to the admissible cut $c$ on the tree $t$. One sees
easily, that $\deg(t) = \deg(P_{c}(t)) + \deg(R_{c}(t))$, for all
admissible cuts $c \in C_t$, and therefore:
$$
 \bar{\Delta}(t) = \sum_{c \in C_t} P_{c}(t) \otimes R_{c}(t) \in
                         \sum_{\substack{p+q= \deg(t),\ p,\,q>0}} \mathcal{H}\grad{p} \otimes \mathcal{H}\grad{q}.
$$
Furthermore, this map is extended by definition to an algebra
morphism on $\mathcal{H}_\mathcal{T}$:
$$
    \Delta\big(\prod_{i=1}^{n}t_i\big):=\prod_{i=1}^{n}\Delta(t_i).
$$

\noindent The best way to get use to this particular coproduct is
to present some examples:
 \allowdisplaybreaks{
\begin{eqnarray*}
 \Delta(\ta1) &=&\;\: \ta1  \otimes 1_\mathcal{T} + 1_\mathcal{T} \otimes \ta1 \notag \\
 \Delta\big(
 \!\!\!\begin{array}{c}
                 \\[-0.5cm]\tb2 \\
               \end{array}\!\! \big) &=&
                         \begin{array}{c}
                           \\[-0.5cm]\tb2 \\
                         \end{array} \!\! \otimes 1_\mathcal{T}
                            + 1_\mathcal{T} \otimes \!\!\!
                         \begin{array}{c}
                           \\[-0.5cm]\tb2 \\
                         \end{array}
                               + \ta1 \otimes \ta1 \label{lad2}\\
 \Delta(\ta1\ta1) &=& \Delta(\ta1)\Delta(\ta1) = \;\: \ta1 \ta1 \otimes 1_\mathcal{T} + 1_\mathcal{T} \otimes
                            \ta1\ta1 + 2\ta1 \otimes \ta1 \\
 \Delta\big(\td31\big) &=& \!\! \begin{array}{c}
                                 \\[-0.5cm]\td31 \\
                              \end{array}\!\! \otimes 1_\mathcal{T}
                               + 1_\mathcal{T} \otimes\!\!\!
                              \begin{array}{c}
                                 \\[-0.5cm]\td31 \\
                              \end{array}\!\! +
                                  2\ta1 \otimes \!\!\!
                                       \begin{array}{c}
                                         \\[-0.5cm]\tb2 \\
                                       \end{array} + \ta1\ta1\otimes \ta1 \label{ex:copTree} \\
 \Delta\big(\ta1\tb2\ \!\big) &=& \Delta\big(\tb2\ta1\ \! \big) = \Delta(\ta1)\Delta\big(\!\!\!\begin{array}{c}
                                                                                         \\[-0.5cm]\tb2 \\
                                                                                        \end{array}\!\! \big) \\
                  &=& \;\: \ta1 \tb2 \otimes 1_\mathcal{T} + 1_\mathcal{T} \otimes \ta1\tb2
                                                            + \ta1 \ta1 \otimes \ta1
                            + \ta1 \otimes \ta1\ta1 +  \tb2 \otimes \ta1 +  \ta1 \otimes \tb2\\
 \Delta\big(\!\!\! \begin{array}{c}
                     \\[-0.5cm] \th43  \\
                   \end{array}\!\!\big)  &=& \!\! \begin{array}{c}
                                                   \\[-0.5cm] \th43 \\
                                                  \end{array}\!\! \otimes 1_\mathcal{T}
                               + 1_\mathcal{T} \otimes\!\!
                                                       \begin{array}{c}
                                                        \\[-0.5cm] \th43 \\
                                                       \end{array}
                               + 3 \ta1 \otimes \!\! \begin{array}{c}
                                                      \\[-0.5cm]\td31 \\
                                                     \end{array}
                               + 3 \ta1\ta1 \otimes\!\! \begin{array}{c}
                                                         \\[-0.5cm]\tb2 \\
                                                        \end{array}
                               + \ta1\ta1\ta1 \otimes \ta1.
\end{eqnarray*}}

\noindent One observes immediately that the vector space $\Cal T$
defines a left coideal, that is~:
$$
    \Cal T \xrightarrow{\Delta} \Cal H_{\Cal T} \otimes \Cal T.
$$
The experienced reader will easily recognize that this coproduct
efficiently stores the so-called wood $\mathfrak{W}(\Gamma)$ for
the Feynman graph $\Gamma$ with its hierarchy of subdivergencies
represented by the tree $t\sim\Gamma$. A wood simply contains all
spinneys\footnote{A spinney is a --possibly non-connected--
subgraph with one-particle irreducible components.} of the
graph~\cite{CaswellKennedy82,EGfields06}. The right-hand side of
above coproduct consists of the cograph following form the
contraction of the corresponding spinney denoted on the left-hand
side.

Connes and Kreimer showed that $\mathcal{H}_{\mathcal{T}}$ with
coproduct (\ref{coprod1}), and counit $\epsilon$ defined by
$\epsilon(1):= 1_k$ and zero else is a connected $\mathbb{Z}_{\geq
0}$-graded commutative, non-cocommutative bialgebra of finite type
and hence a Hopf algebra, with antipode $S$ defined recursively by
$S(1_\mathcal{T})=1_\mathcal{T}$ and:
$$
    S(t):= - t  - \sum_{c \in C_t} S(P_{c}(t))R_{c}(t).
$$
Again, a couple of examples might be helpful here:
 {\allowdisplaybreaks{
\begin{eqnarray}
           S(\ta1) &=& - \ta1 \nonumber\\
           S(\!\!\! \begin{array}{c}
                        \\[-0.4cm] \tb2 \\
                    \end{array}\!\!) &=& - \tb2 - S(\ta1)\ta1
                                      =  - \tb2 + \ta1\ \ta1 \nonumber\\
  S\Big(\!\!\!\! \begin{array}{c}
                        \\[-0.4cm]\td31 \\
                 \end{array}\!\!\!\Big) &=& - \;\td31 - 2S(\ta1) \tb2 -S(\ta1\ta1)\ta1
                                         =  - \td31 + 2 \ta1\ \tb2  - \ta1\ \ta1\ \ta1 \label{cherry}\\
  S\Big(\!\!\!\! \begin{array}{c}
                        \\[-0.4cm]\th43 \\
                 \end{array}\!\!\!\Big)  &=& - \th43
                                             - 3 S(\ta1)\td31
                                             - 3 S(\ta1\ta1)\tb2
                                             - S(\ta1\ta1\ta1)\ta1
                                          =  - \th43
                                             + 3 \ta1\ \td31
                                             - 3 \ta1\ \ta1\ \tb2
                                             + \ta1\ \ta1\ \ta1\ \ta1.
\end{eqnarray}}}

We chose a simple decoration of tree vertices by tree-factorials
\cite{KreimerChen,KreiDelb} defined as follows. Let $t \in
\mathcal{T}$, each primitive cut, that is, each edge $c \in E(t)$,
defines two rooted trees, i.e., the pruned tree $t_{v_c}:=P_c(t)$
and $R_c(t)$. The root of the former is the vertex $v_c \in V(t)$
which had $c$ as its incoming edge. The root of the cotree
$R_c(t)$ is the original root. Let $w(t_{v_c}):=|V(t_{v_c})|$,
then $\deg(t)=\deg(w(t_{v_c}))+\deg(R_c(t))$. The tree-factorial
of $t$ is defined by:
$$
    t^{!}:=\prod_{{c \in C^{(1)}_t \atop |c|=1}} w(t_{v_c})
          =\prod_{v \in V(t)} w(t_v).
$$
The second equality is clear as to each vertex we can associate
the unique incoming edge $c \in E(t)$. As examples we mention:
$$
    \ta1^{!}=1,\; \tb2^{\ \! !}=2,\; \tc3^{\ \! !}=6,\; \td31^{!}=3,\; \tf41^{!}=8,\; \;\makebox{
    and}\;\ \th43^{!}=4.
$$
In the following we decorate each vertex $v \in V(t)$ of a rooted
tree $t \in \Cal T$ by its tree-factorial, $t_v^{!}$. For
notational transparency we omit the decorations on the trees.

Recall the function:
$$
    \int_{0}^{\infty} y^{ - az}(y+c)^{-1-bz} dy =
    B\big((a+b)z,1-az\big)c^{-(a+b)z}\ ,
    \makebox{where}\ B(u,v):= \frac{\Gamma(u)\Gamma(v)}{\Gamma(u+v)},
$$
where $c>0$ and $\Gamma(a)$ is the usual Euler Gamma-function
\cite{KreimerBuch}. We define now the family of functions:
$$
    B_n := B_n(z)
        := B(nz,1 - nz),\qquad n>0.
$$
Following~\cite{KreimerChen,KreiDelb} we define the following
regularized toy-model character $\varphi=\varphi(a,\mu,z)$ from
the Hopf algebra of integer decorated rooted trees $\Cal H_{\Cal
T}$ to $\Cal A := \mathbb{C}[z^{-1},z]][[\log(a/\mu)]]$, $a/\mu
>0$~: $\varphi(1_\mathcal{T})(a,\mu,z) := 1_{\Cal A}$,
\begin{equation} \label{toycharacter}
    \varphi(t)(a,\mu,z) := \left(\frac{a}{\mu}\right)^{-z |t|}
                                     \prod_{v \in V(t)} B_{w(t_{v})},
                                     \; \rm{ for\ }\ t \in \mop{Ker}\epsilon.
\end{equation}
Here, $a$ is assumed to be a dimensional {\sl{external}}
parameter, and $\mu$ is the so-called 't~Hooft mass. The latter is
an arbitrarily chosen parameter of the same dimension as $a$, such
that the ratio $a/\mu$ is a positive number. The parameter $\mu$
introduces an external scale specific to dimensional
regularization~\cite{Collins84}. Let us give a few examples by
applying $\varphi(a,\mu,z)$ to some trees. Defining
$\alpha:=\frac{a}{\mu}$ we find:
 \allowdisplaybreaks{
\begin{eqnarray*}
   \varphi(\ta1)(a,\mu,z) &=& \alpha^{-z}B_1
                           = \alpha^{-z} B(z,1 - z)
                           = \alpha^{-z}\frac{\pi}{\sin\pi z}
                           = \alpha^{-z}\big(\frac{1}{z} + \frac{\pi^2}{6}z + O(z^2)\big)\\
   \varphi(\!\!\! \begin{array}{c}
                        \\[-0.4cm] \tb2 \\
                    \end{array}\!\!) (a,\mu,z)&=&  \alpha^{-2z}B_2 B_1
                                                         = \alpha^{-2z} B(2z,1 - 2z)B(z,1 - z),\\
   \varphi\Big(\!\!\!\! \begin{array}{c}
                          \\[-0.4cm]\td31 \\
                        \end{array}\!\!\!\Big) (a,\mu,z)&=&  \alpha^{-3z}B_3 B^2_1
                                                         = \alpha^{-3z} B(3z,1 - 3z)B(z,1 - z)B(z,1 - z),\\
   \varphi\Big(\!\!\!\! \begin{array}{c}
                         \\[-0.4cm]\th43 \\
                        \end{array}\!\!\!\Big) (a,\mu,z)&=&  \alpha^{-4z}B_4 B^3_1
                                                         = \alpha^{-4z} B(4z,1 - 4z)B(z,1 - z)B(z,1 - z)B(z,1 - z),\\
   \varphi\Big(\!\!\!\! \begin{array}{c}
                         \\[-0.4cm]\tf41 \\
                        \end{array}\!\!\!\Big) (a,\mu,z)&=&  \alpha^{-4z}B_4 B_2 B_1 B_1
                                                         = \alpha^{-4z} B(4z,1 - 4z)B(2z,1 - 2z)B^2(z,1 - z).
\end{eqnarray*}}
These types of models exemplify a rich structure capturing some
aspects of real QFT calculations and we refer the reader to
\cite{BK99,KreimerChen,KreiDelb,KreimerBuch} for more details on
such toy-models.

For latter use we parameterize 't~Hooft's mass:
\begin{equation} \label{parameter}
    \mu \rightarrow \mu(s):=e^{s}\mu, \; s \in \mathbb{R},
\end{equation} such that $\alpha:=a/\mu \to \alpha(s)$, and for a
fixed $a$ we define $\varphi(a,\mu,z) \to
\varphi(\alpha(s),z)=:\varphi(s,z)$.}}\end{exam}


\subsection{Connes--Kreimer's Birkhoff decomposition of Hopf algebra characters}
\label{ssect:CKBirkhoff}

Connes and Kreimer extended the work by
Kreimer~\cite{KreimerHopf,KreimerChen} and established the
connected graded commutative non-cocommutative Hopf algebra of
Feynman graphs corresponding to a perturbative quantum field
theory (pQFT).

Moreover, in the context of minimal subtraction as renormalization
scheme in dimensional regularization Connes and
Kreimer~\cite{CK1,CK2} discovered a unique Birkhoff type
decomposition of Hopf algebra characters with values in the
$\CC$-algebra $\Cal A$ of meromorphic functions capturing the
process of renormalization in pQFT. Namely for any $\varphi \in
G_{\Cal A}$ we have:
\begin{equation}\label{birkhoff1}
    \varphi = \varphi_-^{\star -1} \star \varphi_+,
\end{equation}
where both $\varphi_-$ and $\varphi_+$ belong to $G_{\Cal A}$, and
$\varphi_+(x) \in \Cal A_+$ for any $x \in \Cal H$, whereas
$\varphi_-(x) \in \Cal A_-$ for any $x \in \mop{Ker}\epsilon$.
Here, $\Cal A_-$ is the algebra of polynomials in $(z-z_0)^{-1}$
without constant term (the `pole parts'), and $\Cal A_+$ is the
algebra of meromorphic functions which are holomorphic at $z_0$,
corresponding to the splitting of the $\CC$-algebra $\Cal A=\Cal
A_+ \oplus \Cal A_-$ of meromorphic functions. We denote by $\pi:
\Cal A \to \Cal A_-$ the projection onto $\Cal A_-$.

The components $\varphi_-$ and $\varphi_+$ are given by recursive
formulas. Suppose that $\varphi_-(x)$ and $\varphi_+(x)$ are known
for $x \in \Cal H^{(n-1)}$. Define for $x\in\Cal H^{(n)}$ {\sl
Bogoliubov's preparation map\/}:
\begin{equation}\label{bogoliubov1}
    \bar{R}: x \longmapsto \varphi_{-}\star(\varphi - e)(x)=\varphi(x)+\sum_{(x)}\varphi_-(x')\varphi(x'').
\end{equation}
The components in the Birkhoff decomposition are then given by:
 \allowdisplaybreaks{
\begin{eqnarray}\label{birkhoff2}
  \varphi_-(x)&=&           -\pi \left(\bar{R}(x)\right),\\
  \varphi_+(x)&=&(1_{\Cal A}-\pi)\left(\bar{R}(x)\right).
\end{eqnarray}}

The factor $\varphi_+(x)$ is the {\sl renormalized character\/}
whereas $\varphi_-(x)$ is the sum of {\sl{counter terms}\/} one
must add to $\bar{R}(x)$ to get $\varphi_+(x)$. In the example of
minimal subtraction scheme the {\sl{renormalized value}\/} of the
character $\varphi$ at $z_0 \in \CC$ is the well-defined complex
number $\varphi_+(z_0)$, whereas $\varphi(z_0)$ may not exist. The
fact that $\varphi_-$ and $\varphi_+$ are still characters relies
on the Rota--Baxter property for the projection $\pi$, see e.g.
\cite{E-G-K2,E-G-K3,EGfields06}. Moreover, Connes--Kreimer's
results do not depend on the type of regularization or subtraction
scheme.\smallskip

In terms of the toy-model character
$\varphi(s,z)=\varphi(a,e^{s}\mu,z)$, see equation
(\ref{toycharacter}), with parameterized 't~Hooft mass
(\ref{parameter}), mapping the Hopf algebra $\Cal H_{\Cal T}$ of
non-planar integer decorated rooted trees to $\Cal A :=
\mathbb{C}[z^{-1},z]][[\log(\alpha(s))]]$, $\alpha(s):=a /
(e^{s}\mu) > 0$, which decomposes into:
$$
    \Cal A  = z^{-1}\mathbb{C}[z^{-1}][[\log(\alpha(s))]]
              \oplus
              \mathbb{C}[z]][[\log(\alpha(s))]],
$$
we find for the primitive tree $\ta1 \in \Cal H^{(1)}$ the counter
term:
 \allowdisplaybreaks{
\begin{eqnarray*}
    \varphi_{-}(\ta1)(s,z)&=&-\pi\big(\bar{R}(\ta1)\big)
                      =-\pi\big(\varphi(\ta1)\big)\\
                     &=& -\pi( \alpha(s)^{-z} B_1) \in z^{-1}\mathbb{C}[z^{-1}]
\end{eqnarray*}}
and for $\td31 \in \Cal H^{(3)}$ we find:
 \allowdisplaybreaks{
\begin{eqnarray*}
    \bar{R}\big(\!\!\!\!
                \begin{array}{c}
                 \\[-0.4cm]\td31  \\
                \end{array}\!\!\!\big) &=& \varphi\big(\!\!\!\!
                \begin{array}{c}
                 \\[-0.4cm]\td31  \\
                \end{array}\!\!\!\big) + 2\varphi_{-}(\ta1)\varphi(\!\!\!
                            \begin{array}{c}
                             \\[-0.4cm] \tb2  \\
                            \end{array}\!\!)
                                       + \varphi_{-}(\ta1\ta1)\varphi_{-}(\ta1)\\
                                        &=& \alpha^{-3z}B_3 B^2_1
                                             -2 \pi(\alpha^{-z} B_1)\alpha^{-2z}B_2B_1
                                             +\pi(\alpha^{-2z}B^2_1)\alpha^{-z}B_1,
\end{eqnarray*}}
such that the counter term is given by:
$$
    \varphi_{-}\big(\!\!\!\!
                \begin{array}{c}
                 \\[-0.4cm]\td31  \\
                \end{array}\!\!\!\big)(s,z)
                = - \pi\Big( \varphi\big(\!\!\!\!
                    \begin{array}{c}
                        \\[-0.4cm]\td31  \\
                    \end{array}\!\!\!\big) + 2\varphi_{-}(\ta1)\varphi(\!\!\!
                                \begin{array}{c}
                                 \\[-0.4cm] \tb2  \\
                                \end{array}\!\!)
                    + \varphi_{-}(\ta1\ta1)\varphi_{-}(\ta1)\Big).
$$
A detailed calculation shows that
$\varphi_{-}\big(\!\!\!\!\begin{array}{c}
                            \\[-0.4cm]\td31  \\
                         \end{array}\!\!\!\big)(s,z)
                         \in z^{-1}\mathbb{C}[z^{-1}]$. Hence any
                         dependence on $\mu$ has disappeared.
Feynman rule characters in dimensional regularization depend on
the unit of mass $\mu$. However, for the corresponding counter
terms in their Birkhoff decomposition it is true in general that
$\varphi_{-}(t)(s,z) \in z^{-1}\mathbb{C}[z^{-1}]$, $t \in \Cal
H_{\Cal T}$. In terms of our toy-model character
(\ref{toycharacter}) with parameterized unit mass
$\mu=\mu(s)$~(\ref{parameter}) this may be summarized by saying
that $\partial_s \varphi_{-}(t)(s,z)=0$, $t \in \Cal H_{\Cal T}$.
We will come back to this in section~\ref{sect:matrix}.

\noindent The reader should compare the above counter term with
the expression for
               $S\big(\!\!\!\!
                \begin{array}{c}
                 \\[-0.4cm]\td31  \\
                \end{array}\!\!\!\big)$
in Equation~(\ref{cherry}). Eventually, the corresponding
renormalized expression
              $\varphi_{+}\big(\!\!\!\!
                \begin{array}{c}
                 \\[-0.4cm]\td31  \\
                \end{array}\!\!\!\big) :=(1_{\Cal A}-\pi)\big(\bar{R}
                      \big(\!\!
                        \begin{array}{c}
                            \\[-0.4cm]\td31  \\
                        \end{array}\!\!\!\big) \big)$
is then given by
$$
    \varphi_{-} \star \varphi\big(\!\!\!\!
                \begin{array}{c}
                 \\[-0.4cm]\td31  \\
                \end{array}\!\!\!\big)  =
    \varphi_{-}\big(\!\!\!\!
               \begin{array}{c}
                \\[-0.4cm]\td31  \\
               \end{array}\!\!\!\big)
    + \varphi\big(\!\!\!\!
                \begin{array}{c}
                 \\[-0.4cm]\td31  \\
                \end{array}\!\!\!\big)
    + 2 \varphi_{-}(\ta1)\varphi(\!\!\begin{array}{c}
                                 \\[-0.4cm] \tb2  \\
                                \end{array}\!\!)
    +   \varphi_{-}(\ta1\ta1)\varphi_{-}(\ta1).
$$
The reader should verify that
$\varphi_{-}(\ta1\ta1)=\varphi_{-}(\ta1)\varphi_{-}(\ta1)$, that
is, $\varphi_{-} \in G_{\Cal A}$, hence $\varphi_{+}=\varphi_{-}
\star \varphi \in G_{\Cal A}$.


\subsection{The matrix representation}
\label{ssect:matrix}

In this section we recall the matrix representation of $\Cal
L(\Cal H,\Cal A)$ associated with a left coideal. The next step
consists in understanding the beta-function of
Connes--Kreimer~\cite{CK3} in this matrix setting. We follow
Sakakibara's approach~\cite{Sa04}, giving his clever computations
the concrete support of triangular matrices.\\

Let us start by retrieving some material from \cite
{EG05,EGfields06,EGMbch06}. Recall that the subalgebra
$\mathcal{M}^\ell_n(\Cal A) \subset \mathcal{M}_n(\Cal A)$ of
lower triangular matrices in the algebra of $n \times n$ matrices
with entries in the algebra $\Cal A$, and with $n$ finite or
infinite has a decreasing filtration and is complete in the
induced topology. Indeed, $\mathcal{M}^\ell_n(\Cal A)^{m}$ is the
ideal of strictly lower triangular matrices with zero on the main
diagonal and on the first $m-1$ subdiagonals, $m>1$. We then have
the decreasing filtration
$$
   \mathcal{M}^\ell_n(\Cal A) \supset \mathcal{M}^\ell_n(\Cal A)^{1} \supset \dots
  \supset \mathcal{M}^\ell_n(\Cal A)^{m-1} \supset \mathcal{M}^\ell_n(\Cal A)^{m} \supset \cdots, m<n,
$$
with
$$
  \mathcal{M}^\ell_n(\Cal A)^{u}\: \mathcal{M}^\ell_n(\Cal A)^{v} \subset
  \mathcal{M}^\ell_n(\Cal A)^{u+v}.
$$
For $\Cal A$ being commutative we denote by $\mathfrak{M}_n(\Cal
A)$ the group of lower triangular matrices with unit diagonal
which is $\mathfrak{M}_n(\Cal A) = {\bf{1}} +
\mathcal{M}^\ell_n(\Cal A)^{1}$. Here the $n \times n$, $n\leq
\infty$, unit matrix is given by
\begin{equation}
    \label{def:unitmatrix}
     {\bf 1}:=(\delta_{ij}1_\Cal A)_{1\leq i,j\leq n }.\\
\end{equation}

Let $\Cal H$ be a connected filtered Hopf algebra over $k$, let
$\Cal A$ be any commutative unital $k$-algebra, and let $\big(\Cal
L(\Cal H,\Cal A),\star\big)$ be the algebra of $k$-linear maps
from $\Cal H$ to $\Cal A$ endowed with the convolution product.
Let $J$ be any left coideal of $\Cal H$ (i.e. a vector subspace of
$\Cal H$ such that $\Delta(J)\subset \Cal H\otimes J$).

We fix a basis $X=(x_i)_{i\in I}$ of the left coideal $J$ (a {\sl
left subcoset\/} in the terminology of \cite{EG05}). We suppose
further that this basis is denumerable (hence indexed by $I=\NN$
or $I=\{1,\cdots ,m\})$ and {\sl filtration ordered\/}, i.e. such
that if $i\le j$ and $x_j\in\Cal H^{(n)}$, then $x_i\in\Cal
H^{(n)}$.

\begin{defn} \label{def:coproductmatrix}
The {\sl coproduct matrix\/} in the basis $X$ is the $|I|\times |I|$ matrix
$M$ with entries in $\Cal H$ defined by~:
$$
    \Delta(x_i)=\sum_{j\in I}M_{ij}\otimes x_j.
$$
\end{defn}

\begin{lem}
The coproduct matrix is lower-triangular with diagonal terms equal
to $1$.
\end{lem}

\begin{proof}
Suppose $x_i \in \Cal H^{(n)}$ and $x_i\notin\Cal H^{(n-1)}$. Then
it is well-known that (see e.g. \cite{FG05}, \cite{Ma01}):
$$
    \Delta(x_i) = x_ i \otimes 1 + 1 \otimes x_i + \hbox{ terms of filtration degree }\le n-1.
$$
Then clearly $M_{ii}=1$, and moreover if $M_{ij}\not =0$ and $i\not =j$, then $x_j\in\Cal
H^{(n-1)}$. If $i<j$ this implies $x_i\in H^{(n-1)}$, which contradicts the
hypothesis. Hence $M_{ij}=0$ if $i<j$.
\end{proof}

Recalling the example of rooted trees and choosing the following
subset $\Cal T' \subset \Cal T$ of rooted trees with the displayed
linear order:
\begin{equation}
\mathcal{T}':=
 \left\{
t_1:= 1_\mathcal{T},\
t_2:= \ta1\, ,\
t_3:=\!\! \begin{array}{c} \tb2 \\ \end{array}\!\!,\
t_4:=\!\!\!\begin{array}{c} \td31\ \\ \end{array}\!\!\!,\
t_5:=\!\!\!\begin{array}{c} \th43\ \\ \end{array}\!\!\! \right\}.
\label{6space}
\end{equation}
The $5 \times 5$ coproduct matrix of
Definition~\ref{def:coproductmatrix} is then given by:
\begin{equation}
 \label{coprodMat6}
 M = \left(
  \begin{array}{cccccc}
   1_\mathcal{T}   & 0                & 0               & 0             &0 \\
   \at1            & 1_\mathcal{T}    & 0               & 0             &0 \\
   \mat41          & \ta1             & 1_\mathcal{T}   & 0             &0 \\
   \Mat31          & \ta1\ta1         & 2 \ta1          & 1_\mathcal{T} &0 \\
   \mot43          & \ta1\ta1\ta1     & 3 \ta1\ta1      & 3\ta1         & 1_\mathcal{T}
  \end{array}
           \right)
\end{equation}
Observe that for each tree $t_i \in \Cal T'$ all cotrees in
$\bar{\Delta}(t_i)$ are of degree strictly lower than $\deg(t_i)$
and are contained in $\Cal T'$, hence $\Cal T'$ forms a left
coideal in $\Cal H_{\Cal T}$.

Now define $\Psi_J:\Cal L(\Cal H,\Cal A) \to \mop{End}_{\Cal
A}(\Cal A\otimes J)$ by~:
\begin{equation}\label{Psi}
    \Psi_J[f](x_j)=\sum_if(M_{ij})\otimes x_i.
\end{equation}
In other words, the matrix of $\Psi_J[f]$ is given by
$\big(f(M_{ij})\big)_{i,j\in I}$.

\begin{prop}\label{morphism}
The map $\Psi_J$ defined above is an algebra homomorphism. Its
transpose does not depend on the choice of the basis.
\end{prop}

\begin{proof}
The second statement is straightforward: let us denote by
${}^{\top}\!\Psi_J$ the transpose of $\Psi_J$, i.e. the map
defined by:
$$
    {}^{\top}\!{\Psi_J}[f](x_i) = \sum_j f(M_{ij}) \otimes x_j.
$$
For any $x \in J$ we have then, using Sweedler's notation:
$$
    {}^{\top}\!{\Psi_J}[f](x) = \sum_{(x)} f(x_{(1)})\otimes x_{(2)}.
$$
Hence ${}^{\top}\!{\Psi_J}[f]$ has an intrinsic expression as the
composition of the three maps below:
$$
    \Cal A\otimes J\fleche{14}^{Id_{\Cal A} \otimes \Delta}
    \Cal A\otimes\Cal H\otimes J\fleche{22}^{Id_{\Cal A} \otimes f \otimes Id_J}
    \Cal A\otimes\Cal A\otimes J\fleche{16}^{m_{\Cal A} \otimes Id_J}
    \Cal A\otimes J.
$$
We have to show for any $f,g \in \Cal L(\Cal H,\Cal A)$:
\begin{equation}\label{morphism2}
    \Psi_J[f \star g] = \Psi_J[f]\Psi_J[g].
\end{equation}
It is shown in \cite{EG05} that ${}^{\top}\!\Psi_J$ is an
anti-homomorphism, which proves the claim. We give here an
alternative proof: Using coassociativity $(Id \otimes \Delta)
\circ \Delta(x_i) = (\Delta \otimes Id) \circ \Delta(x_i)$, we
immediately get~:
$$
    \Delta(M_{ij}) = \sum_{k=0}^{|I|}M_{ik}\otimes M_{kj}.
$$
Hence,
 \allowdisplaybreaks{
\begin{eqnarray*}
    \Psi_J[f \star g](x_j)&=& \sum_i(f \star g)(M_{ij})\otimes x_i\\
                    &=& \sum_i\sum_kf(M_{ik})g(M_{kj})\otimes x_i\\
                    &=& \Psi_J[f]\circ\Psi_J[g](x_j),
\end{eqnarray*}}
which proves (\ref{morphism2}).
\end{proof}

The Lie algebra of infinitesimal characters is mapped into a Lie
subalgebra $\widehat{\g g}_{\Cal A}$ in $\Cal M^\ell_{|I|}(\Cal
A)^1$. This is immediately seen by applying definition
(\ref{Psi}), since elements in $\g g_A$ map the unit to zero due
to relation (\ref{infinitesimal}). Whereas characters in the group
$G_{\Cal A}$ are mapped to the subgroup $\widehat{G}_{\Cal A}
\subset \mathfrak{M}_{|I|}(\Cal A)$.

The toy-model character $\varphi=\varphi(s,z)$ in
equation~(\ref{toycharacter}) with parameterized 't~Hooft mass,
applied to the coproduct matrix (\ref{coprodMat6}) gives
\begin{equation*}
 \widehat{\varphi}(s,z) = {\small{\left(
  \begin{array}{cccccc}
   1_{\Cal A}      & 0            & 0              & 0            &0 \\
   \alpha(s)^{-z}B_1       & 1_{\Cal A}   & 0              & 0            &0 \\
   \alpha(s)^{-2z}B_2B_1   & \alpha(s)^{-z}B_1    & 1_{\Cal A}     & 0            &0 \\
   \alpha(s)^{-3z}B_3B^1_1 & \alpha(s)^{-2z}B^2_1 & 2 \alpha(s)^{-z}B_1    & 1_{\Cal A}   &0 \\
   \alpha(s)^{-4z}B_4B^3_1 & \alpha(s)^{-3z}B^3_1 & 3 \alpha(s)^{-2z}B^2_1 & 3\alpha(s)^{-z}B_1   & 1_{\Cal A}
  \end{array}
           \right)}}
\end{equation*}
Recall that $\alpha=\alpha(s)=a/\mu(s)$, where $\mu(s)=e^{s}\mu$.

The following remarks should be useful latter. The coproduct
matrix $M$ with entries in $\Cal H$ can be seen as the image of
the identity map under  $\Psi_J:\Cal L(\Cal H,\Cal H) \to
\mop{End}_{\Cal H}(\Cal H \otimes J)$, i.e.~:
\begin{equation}\label{PsiId}
    \Psi_J[Id](x_j) = \sum_i Id(M_{ij})\otimes x_i.
\end{equation}
The equations (\ref{antipode}) imply for the matrix representation
of the antipode $S \in \Cal L(\Cal H,\Cal H)$~:
 \allowdisplaybreaks{
\begin{eqnarray*}
    \Psi_J[S \star Id](x_j) &=& \Psi_J[S]\circ\Psi_J[Id](x_j)\\
                            &=& \sum_i\sum_k S(M_{ik})Id(M_{kj})\otimes x_i\\
                            &=& \Psi_J[\eta \circ \epsilon](x_j)\\
                            &=& \sum_i \eta \circ \epsilon(M_{ij})\otimes x_i\\
                            &=& \sum_i {\bf{1}}\delta_{ij} \otimes x_i.
\end{eqnarray*}}
Here ${\bf{1}}$ denotes the $|I|\times |I|$ unit matrix,
$(\delta_{ij}1_\Cal T)_{1 \leq i,j\leq |I|}$. Hence, $\Psi_J[S] =
M^{-1}$, and the inverse can be calculated readily in terms of the
geometric series:
$$
    \Psi_J[S]= M^{-1}= {\bf{1}} + \sum_{k>0} (-1)^k (M-{\bf{1}})^k.
$$
Using the bijection $\exp^{\star}$ between $\g g_{\Cal A}$ and
$G_{\Cal A}$ we may write any $\varphi \in G_{\Cal A}$ as an
exponential of the element $\alpha=\log^{\star}(\varphi)$ in $\g
g_{\Cal A}$. In terms of matrices we find $\log(\Psi[\varphi]) =
\varphi(\log(\Psi[Id]))$, using the fact that $\varphi$ is a
character. Such that:
$$
    \log(\Psi[Id]) = \log(M)= \sum_{k>0} (-1)^k \frac{(M-{\bf{1}})^k}{k}
$$
defines the matrix of the so-called {\it{normal coordinates}\/}.


\subsection{The matrix form of Connes--Kreimer's Birkhoff decomposition}
\label{subsect:matrixCK}

Suppose that the commutative target space algebra $\Cal A$ in
$\Cal L(\Cal H, \Cal A)$ splits into two subalgebras:
\begin{equation}\label{split}
    \Cal A = \Cal A_- \oplus \Cal A_+,
\end{equation}
where the unit $1_{\Cal A}$ belongs to $\Cal A_+$. Let us denote
by $\pi: \Cal A \to \Cal A_-$ the projection onto $\Cal A_-$
parallel to $\Cal A_+$. One readily verifies that $\pi$ is an
idempotent {\sl{Rota--Baxter operator}\/}~\cite{EGfields06}, that
is, it satisfies the relation:
\begin{equation} \label{RBR}
    \pi\big( \pi(a)b + a\pi(b) - ab \big) = \pi(a)\pi(b).
\end{equation}
Indeed, let $a,b \in \Cal A$~:
\begin{align*}
    \pi(a)b + a\pi(b) - ab &= \pi(a)\pi(b) - (Id_{\Cal A}-\pi)(a)(Id_{\Cal A}-\pi)(b)
\end{align*}
such that applying $\pi$ on both sides gives relation (\ref{RBR}),
since it eliminates the term $(Id_{\Cal A}-\pi)(a)(Id_{\Cal
A}-\pi)(b)$ without changing the term $\pi(a)\pi(b)$, as $\pi
(Id_{\Cal A}-\pi)(a)=0$ and $\Cal A_- = \pi(\Cal A)$, $\Cal
A_+=(Id_{\Cal A}-\pi)(\Cal A)$ are subalgebras.

In fact, this is a special case of an additive decomposition
theorem characterizing Rota--Baxter algebras which was proven by
Atkinson for general, not necessarily associative algebras in
\cite{Atkinson}.

\begin{thm} \label{Atkinson1} {\rm{(Atkinson, \cite{Atkinson})}}
Let $\Cal A$ be a $k$-algebra. A $k$-linear operator $R :\Cal A
\to \Cal A$ satisfies the Rota--Baxter relation (\ref{RBR}) if and
only if the following two statements are true. Firstly, $\Cal
A_+:=R(\Cal A) $ and $\Cal A_-:=\tilde{R}(\Cal A)$ are subalgebras
in $\Cal A$. Secondly, for $x,y,z \in \Cal A$, $R(x)R(y)=R(z)$
implies $\tilde{R}(x)\tilde{R}(y) = -\tilde{R}(z)$. Here we
denoted the map $\tilde{R}:=(Id_{\Cal A}-R)$.
\end{thm}

The case of an idempotent Rota--Baxter map implies $\Cal A_{-}
\cap \Cal A_{+} = \{0\}$. In the context of perturbative
renormalization in QFT where the regularization prescription
implies the target space algebra $\Cal A$, the corresponding
splitting of $\Cal A$ into a direct sum of two subalgebras is
called a {\sl{renormalization scheme}\/}. For example, the
{\sl{minimal renormalization scheme}\/} in {\sl{dimensional
regularization}\/} corresponds to the splitting of the
$\CC$-algebra $\Cal A$ of meromorphic functions in which $\Cal
A_-$ is the algebra of polynomials in $(z-z_0)^{-1}$ without
constant term (the ``pole parts''), and $\Cal A_+$ is the algebra
of meromorphic functions which are holomorphic at $z_0$.
\medskip

The following theorem describes a multiplicative decomposition for
associative unital Rota--Baxter algebras and was observed by
Atkinson in~\cite{Atkinson}, see also \cite{EGMbch06}.

\begin{thm} \label{Atkinson2} {\rm{(Atkinson, \cite{Atkinson})}}
Let $\Cal A$ be an associative unital Rota--Baxter algebra with
Rota--Baxter map $R$. Suppose $\Cal A$ to be have a decreasing
filtration and to be complete in the induced topology. Assume $X$
and $Y$ in $\Cal A$ to be solutions of the equations:
\begin{equation} \label{MatrixBogo}
        X = 1_{\Cal A} - R(X\ a) \ {\rm{and}}\  Y = 1_{\Cal A} - \tilde{R}(a\ Y),
\end{equation}
for $a \in \Cal A^{(1)}$. Then we have the following factorization
\begin{equation}
        X (1_{\Cal A} + a) Y = 1_{\Cal A},\ \makebox{ such that }\ 1_{\Cal A} + a= X^{-1}Y^{-1}.
    \label{eq:Atkinsonfact1}
\end{equation}
For an idempotent Rota--Baxter map this factorization is unique.
\end{thm}

\begin{proof}
The Rota--Baxter relation~(\ref{RBR})~yields for any
$\alpha,\beta\in\Cal A$~:
\begin{equation}\label{RBRbis}
R\big(\alpha\tilde R(\beta)\big)+\tilde R\big(R(\alpha)\beta\big)
=R(\alpha)\tilde R(\beta).
\end{equation}
We then simply calculate the product $XY$ and use equation
(\ref{RBRbis}), with $\alpha=Xa$ and $\beta=aY$~:
 \allowdisplaybreaks{
\begin{eqnarray*}
        XY &=& \big(1_{\Cal A}-R(X\ a) \big)\ \big(1_{\Cal A}-\tilde{R}(a\ Y)\big)                  \nonumber\\
            &=& 1_{\Cal A} - R(X\ a) - \tilde{R}(a\ Y) + R(X\ a)\tilde{R}(a\ Y)            \nonumber\\
            &=& 1_{\Cal A} - R\Big(X a\ \big(1_{\Cal A}-\tilde{R}(a\ Y)\big)\Big)
                                            - \tilde{R}\Big(\big(1_{\Cal A}-R(X\ a)\big)\ a\ Y\Big)  \\
            &=& 1_{\Cal A} - R(XaY) - \tilde{R}(XaY)                                       \nonumber\\
            &=& 1_{\Cal A} - XaY.
\end{eqnarray*}}
Hence, we obtain the factorization in (\ref{eq:Atkinsonfact1}).
The uniqueness for an idempotent Rota--Baxter map is easy to
show~\cite{E-G-K3,EGMbch06}.
\end{proof}

Let us come back to the matrix representation of $\Cal L(\Cal H,
\Cal A)$ with a splitting $\Cal A$~(\ref{split}) via
$\Psi_{J}$~(\ref{Psi}). We define a Rota--Baxter map $\mathrm{R}$
on $\Psi_{J}[\Cal L] \subset \mathcal{M}^\ell_{|I|}(\Cal A)$ by
extending the Rota--Baxter map $\pi$ on $\Cal A$ entrywise, i.e.,
for the matrix $\tau =(\tau_{ij}) \in \mathcal{M}^\ell_{|I|}(\Cal
A)$, define:
\begin{equation}
    \label{RBonMatrices}
    \mathrm{R}(\tau) =  \big( \pi(\tau_{ij}) \big).
\end{equation}

\begin{thm} \label{thm:RBalgMatrices} {\rm{\cite{EG05}}}
Then the triple $\left(\mathcal{M}^\ell_{|I|}(\Cal A), \mathrm{R},
\{\mathcal{M}^\ell_{|I|}(\Cal A)^l\}_{l<|I|} \right)$ forms a
non-commutative complete filtered Rota--Baxter algebra with
idempotent Rota--Baxter map $\mathrm{R}$.
\end{thm}

Atkinson's multiplicative decomposition immediately implies a
factorization of $\widehat{G}_{\Cal A} \subset
\mathfrak{M}^\ell_{|I|}(\Cal A)$ into the subgroups:
$$
    \widehat{G}^{-}_{\Cal A} \subset {\bf{1}}+\mathrm{R}\big(\Cal M_{|I|}^\ell({\Cal A})^1\big)
                             \subset \mathfrak{M}^\ell_{|I|}(\Cal A)
$$
and
$$
    \widehat{G}^{+}_{\Cal A} \subset {\bf{1}}+\tilde{\mathrm{R}}\big(\Cal M_{|I|}^\ell({\Cal A})^1\big)
                             \subset \mathfrak{M}^\ell_{|I|}(\Cal A),
$$
that is, for each $\widehat{\varphi}:=\Psi[\varphi] \in
\widehat{G}_{\Cal A}$, $\varphi \in G_{\Cal A}$ there exist unique
$\widehat{\varphi}_- \in \widehat{G}^{-}_{\Cal A}$ and
$\widehat{\varphi}_+ \in \widehat{G}^{+}_{\Cal A}$, such that:
\begin{equation}\label{matrixBirkhoff}
    \widehat{\varphi} = \widehat{\varphi}{}^{-1}_-\ \widehat{\varphi}_+.
\end{equation}
We immediately see that $\widehat{\varphi}{}_-$ and
$\widehat{\varphi}^{-1}_+$ are unique solutions of the equations
(\ref{MatrixBogo}) in Theorem~\ref{Atkinson2}:
 \allowdisplaybreaks{
\begin{eqnarray}
    \widehat{\varphi}_-  &=& {\bf{1}} - \mathrm{R}\Big(\widehat{\varphi}_- \ (\widehat{\varphi}-{\bf{1}})\Big),
    \label{eq:mphi-}\\
    \widehat{\varphi}^{-1}_+  &=& {\bf{1}} - \tilde{\mathrm{R}}\Big( (\widehat{\varphi}-{\bf{1}})\ \widehat{\varphi}^{-1}_+ \Big).
    \label{eq:mphi+}
\end{eqnarray}}
Moreover, after some simple algebra using the factorization
$\widehat\varphi=\widehat\varphi_-^{-1}\widehat\varphi_{+}$~:
$$
    \widehat{\varphi}_+ (\widehat{\varphi}^{-1}-{\bf{1}})\ = \widehat{\varphi}_{-} - \widehat{\varphi}_+
                                                           = - \widehat{\varphi}_{-}(\widehat{\varphi}-{\bf{1}})
$$
we immediately get the recursion for
$\widehat{\varphi}_+$~\cite{EG05}:
\begin{equation}\label{eq:mmphi+}
    \widehat{\varphi}_+  = {\bf{1}} - \tilde{\mathrm{R}}\Big(\widehat{\varphi}_+ \ (\widehat{\varphi}^{-1}-{\bf{1}}) \Big).
\end{equation}
and hence we see that
\begin{equation}\label{eq:mmmphi+}
    \widehat{\varphi}_+  = {\bf{1}} + \tilde{\mathrm{R}}\Big(\widehat{\varphi}_- \ (\widehat{\varphi}-{\bf{1}}) \Big).
\end{equation}

\noindent The matrix entries of $\widehat{\varphi}_-$ and
$\widehat{\varphi}^{-1}_+$ can be calculated without recursions
using $\sigma := \widehat{\varphi}$ from the equations:
 \allowdisplaybreaks{
\begin{eqnarray*}
    (\widehat{\varphi}_-)_{ij} &\!=\!& -\pi(\sigma_{ij})\!
                                - \sum_{k=2}^{j-i}\: \sum_{i>l_1>l_2> \cdots >l_{k-1}>j} (-1)^{k+1}
       \pi\big(\pi(\cdots \pi(\sigma_{i l_{1}})\sigma_{l_{1} l_{2}}) \cdots \sigma_{ l_{k-1} j}\big)\\
    (\widehat{\varphi}^{-1}_+)_{ij} \!&\!\!=\!\!&\!\!  -\tilde{\pi}((\sigma^{-1})_{ij})
                                \!-\!\!\sum_{k=2}^{j-i}\: \sum_{i>l_1>l_2>\cdots > l_{k-1} > j}\!\! (-1)^{k+1}
        \tilde{\pi}\big(\tilde{\pi}(\cdots \tilde{\pi}
        ((\sigma^{-1})_{i l_{1}})(\sigma^{-1})_{l_{1} l_{2}})\cdots (\sigma^{-1})_{ l_{k-1}
        j}\big),
\end{eqnarray*}}
where $\tilde{\pi}:=1_\Cal{A}-\pi$. The matrix entries of
$\widehat{\varphi}_+$ follow from the first formula, i.e., the one
for the entries in $\widehat{\varphi}_-$, by replacing $\pi$ by
$-\tilde{\pi}$. We may therefore define the matrix:
 \allowdisplaybreaks{
\begin{eqnarray} \label{BogoRbar}
 \widehat{{\bar{R}}}[\varphi]:=\widehat{\varphi}_- \ (\widehat{\varphi}-{\bf{1}})
\end{eqnarray}}
such that:
 \allowdisplaybreaks{
\begin{eqnarray} \label{matrixAtkinson}
 \widehat{\varphi}_- = {\bf{1}} - \mathrm{R}\Big(\widehat{\bar{R}}[ \varphi]\Big)\
 \makebox{ and }\
 \widehat{\varphi}_+ = {\bf{1}} + \tilde{\mathrm{R}}\Big(\widehat{\bar{R}}[\varphi]\Big).
\end{eqnarray}}
In fact, equations (\ref{eq:mmmphi+}) and (\ref{eq:mphi+}) may be
called {\it{Bogoliubov's matrix formulae}\/} for the counter term
and renormalized Feynman rules matrix, $ \widehat{\varphi}_-$, $
\widehat{\varphi}_+$, respectively. Equation~(\ref{BogoRbar}) is
the matrix form of Bogoliubov's $\bar{R}$- or preparation
map~(\ref{bogoliubov1}), e.g. see~\cite{Collins84}~:
\begin{equation}
 \label{bogoliubov2}
    \widehat{{\bar{R}}}[\varphi] := \Psi_{J}[\bar{R}]
                                  = \Psi_{J}[\varphi_{-}\star(\varphi - e)].
\end{equation}
Here, $\varphi_{-}$ is the counter term
character~(\ref{birkhoff2}) of the algebraic {\sl{Birkhoff
decomposition}\/} of Connes and Kreimer \cite{CK1,CK2}, which we
mentioned in the foregoing section. To phrase it differently, the
above matrix factorization follows via the representation
$\Psi_{J}$ from Connes--Kreimer's Birkhoff
factorization~(\ref{birkhoff1}) on the group $G_{\Cal A}$ of $\Cal
A$-valued Hopf algebra characters.

\begin{rmk} \label{bch-CHI} {\rm{We may apply the result from
\cite{E-G-K2,E-G-K3,EG05,EGMbch06}, see also \cite{Ma01}, to the
above matrix representation of $\g g_{\Cal A}$ respectively
$G_{\Cal A}$. In these references a unique non-linear map $\chi$
was established on $\g g_{\Cal A}$ which allows to write the
characters $\varphi_-$ and $\varphi_+$ as exponentials. In the
matrix picture we hence find for $\widehat{Z} \in  \widehat{\g
g}_{\Cal A}$ and  $\widehat{\varphi}=\exp(\widehat{Z}) \in
\widehat{G}_{\Cal A}$~:
\begin{equation}
\label{bch}
    \widehat{\varphi}=\exp\big(\mathrm{R}(\chi(\widehat{Z}))\big)
                      \exp\big(\tilde{\mathrm{R}}(\chi(\widehat{Z}))\big).
\end{equation}
The matrices
$\widehat{\varphi}_-:=\exp\big(-\mathrm{R}(\chi(\widehat{Z}))\big)$
and
$\widehat{\varphi}^{-1}_+:=\exp\big(-\tilde{\mathrm{R}}(\chi(\widehat{Z}))\big)$
are in $\widehat{G}^{-}_{\Cal A}$ and $\widehat{G}^{+}_{\Cal A}$,
respectively, and solve Bogoliubov's matrix formulae in
(\ref{matrixAtkinson}). }}\end{rmk}


\section{The matrix representation of the beta-function}
\label{sect:matrix}

It is the goal of this section to establish a matrix
representation of the beta-function as it appears in the work of
Connes and Kreimer~\cite{CK3}. Once we have achieved this we
reformulate in a transparent manner Sakakiabara's
findings~\cite{Sa04} hereby providing a firm ground for his
calculations. In the next paragraph we review the main points of
the beta-function calculus in the Hopf algebra context following
mainly the paper~\cite{Ma01}.


\subsection{The beta-function in the Hopf algebra of renormalization}
\label{subsect:beta}

From now on, $k=\CC$ stands for the complex numbers, and $\Cal A$
will denote the algebra of meromorphic functions in one complex
variable $z$ endowed with the minimal subtraction scheme at
$z_0=0$. Hence $\Cal A=\Cal A_-\oplus\Cal A_+$, where $\Cal A_+$
is the subalgebra of meromorphic functions which are holomorphic
at $0$, and $\Cal A_-$ stands for the polynomials in $z^{-1}$
without constant term. We moreover suppose that the Hopf algebra
$\Cal H$ is graded, with filtration coming from the graduation,
i.e.~:
$$
    \Cal H^{(n)} = \bigoplus_{0\le k\le n} \Cal H_k.
$$
The grading induces a biderivation $Y$ defined on homogeneous
elements by~:
 \allowdisplaybreaks{
\begin{eqnarray*}
    Y:\Cal H_n  &\longrightarrow \Cal H_n   \\
            x   &\longmapsto    nx.
\end{eqnarray*}}
Exponentiating $Y$ we get a one-parameter group $\theta_t$ of
automorphisms of the Hopf algebra $\Cal H$, defined on $\Cal H_n$
by~:
\begin{equation} \label{group}
    \theta_t(x)=e^{nt}x.
\end{equation}
The map $\varphi \mapsto \varphi \circ Y$ is a derivation of
$\big(\Cal L(\Cal H,\Cal A),\star \big)$, and $\varphi \mapsto
\varphi \circ \theta_t$ is an automorphism of $\big(\Cal L(\Cal
H,\Cal A),\star\big)$ for any complex $t$. We will rather consider
the one-parameter group $\varphi \mapsto \varphi \circ
\theta_{tz}$ of automorphisms of the algebra  $\big(\Cal L(\Cal
H,\Cal A),\star\big)$ i.e.~:
\begin{equation}
    \varphi^t(x)(z):=e^{tz|x|}\varphi(x)(z).
\end{equation}
Differentiating at $t=0$ we get:
\begin{equation}\label{gen}
    {d \over dt}\restr{t=0} \varphi^t = z(\varphi \circ Y).
\end{equation}

We denote by $G^\Phi_{\Cal A}$ the set of those characters
$\varphi \in G$ such that the negative part of the Birkhoff
decomposition of $\varphi^t$ does not depend on $t$, namely:
$$
    G_{\Cal A}^{\Phi}=\Big\{\varphi\in G_{\Cal A} \Big{|}\ {d \over dt}(\varphi^t)_-=0\Big\}.
$$
In particular the dimensional-regularized Feynman rules verify
this property: in physical terms, the counter terms do not depend
on the choice of the arbitrary mass-parameter $\mu$ ('tHooft's
mass) one must introduce in dimensional regularization in order to
get dimensionless expressions (see \cite {CK2}). We also denote by
$G_{\Cal A_{-}}^\Phi$ the elements $\varphi$ of $G_{\Cal A}^ \Phi$
such that $\varphi = \varphi_-^{\star -1}$. Recall from
\cite{Ma01} that there is a bijection $\wt R: G_{\Cal A} \to \g
g_{\Cal A}$ defined by:
\begin{equation}\label{E}
    \varphi \circ Y = \varphi \star \wt R(\varphi).
\end{equation}
Since composition on the right with $Y$ is a derivation for the
convolution product, the map $\wt R$ verifies a cocycle property:
\begin{equation}\label{cocycle}
    \wt R(\varphi\star\psi)=\wt R(\psi)+\psi^{\star-1} \star \wt R(\varphi)\star\psi.
\end{equation}
We summarize some key results of \cite{CK3} in the following
proposition:
\goodbreak
\begin{prop}\label{ck2}
\
\begin{enumerate}

\item \label{it:one} For any $\varphi\in G_{\Cal A}$ there is a
one-parameter family $h_t$ in  $G_{\Cal A}$ such that
$\varphi^t=\varphi\star h_t$, and we have:
\begin{equation}\label{gen2}
    \dot h_t := {d \over dt} h_t = h_t \star z\wt R(h_t) + z\wt R(\varphi)\star h_t.
\end{equation}

\item \label{it:two} $z \wt R$ restricts to a bijection from
$G_{\Cal A}^\Phi$ onto $\g g_{\Cal A} \cap \Cal L(\Cal H,\Cal
A_+)$. Moreover it is a bijection from $G_{\Cal A\ -}^\Phi$ onto
those elements of
  $\g g_{\Cal A}$ with values in the constants, i.e.~:
$$
    \g g_{\Cal A}^c= \g g_{\Cal A}\cap \Cal L(\Cal H,\CC).
$$

\item \label{it:three} For $\varphi\in G_{\Cal A}^\Phi$, the
constant term of $h_t$, defined by:
\begin{equation}\label{rg}
    F_t(x) = \mopl{lim}_{z\to 0}h_t(x)(z)
\end{equation}
is a one-parameter subgroup of  $G_{\Cal A} \cap \Cal L(\Cal
H,\CC)$, the scalar-valued characters of $\Cal H$.
\end{enumerate}
\end{prop}

\begin{proof}
For any $\varphi\in G_{\Cal A}$ one can write:
\begin{equation}\label{psi-t}
    \varphi^t = \varphi \star h_t
\end{equation}
with $h_t\in G_{\Cal A}$.  From (\ref{psi-t}), (\ref{gen}) and
(\ref{E}) we immediately get:
$$
    \varphi \star \dot h_t = \varphi \star h_t \star z \wt R(\varphi \star  h_t).
$$
Equation (\ref{gen2}) then follows from the cocycle property
(\ref{cocycle}). This proves the first assertion.

Now take any character $\varphi \in G_{\Cal A}^\Phi$ with Birkhoff
decomposition $\varphi=\varphi_{-}^{{\star}-1} \star \varphi_+$
and write the Birkhoff decomposition of $\varphi^t$~:
 \allowdisplaybreaks{
\begin{eqnarray*}
    \varphi^t &=& ({\varphi^t})_{-}^{{\star}-1} \star (\varphi^t)_+\\
              &=& (\varphi_-)^{{\star}-1}  \star  (\varphi^t)_+\\
              &=& (\varphi \star \varphi_+^{{\star}-1}) \star (\varphi^t)_+\\
              &=& \varphi \star  h_t,
\end{eqnarray*}}
with $h_t$ taking values in $\Cal A_+$. Then $z\wt R(\varphi)$
also takes values in $\Cal A_+$, as a consequence of equation
(\ref{gen2}) at $t=0$. Conversely, suppose that $z\wt R(\varphi)$
takes values in $\Cal A_+$. We show that $h_t$ also takes values
in $\Cal A_+$ for any $t$, which immediately implies that
$\varphi$ belongs to $G_{\Cal A}^\Phi$.

For any $\gamma\in\g g_{\Cal A}$, let us introduce the linear
transformation $U_\gamma$ of $\g g_{\Cal A}$ defined by~:
$$
    U_\gamma(\delta) := \gamma \star \delta + z\delta \circ Y.
$$
If $\gamma$ belongs to $\g g_{\Cal A} \cap\Cal L(\Cal H,\Cal A_+)$
then $U_\gamma$ restricts to a linear transformation of $\g
g_{\Cal A} \cap\Cal L(\Cal H,\Cal A_+)$.

\begin{lem}\label{fix}
For any $\varphi \in G_{\Cal A}$, $n \in \NN$ we have~:
$$
    z^n \varphi \circ Y^{n} = \varphi \star U_{z\wt R(\varphi)}^n(e).
$$
\end{lem}

\begin{proof}
Case $n=0$ is obvious, $n=1$ is just the definition of $\wt R$. We
check thus by induction, using again the fact that composition on
the right with $Y$ is a derivation for the convolution product~:
 \allowdisplaybreaks{
\begin{eqnarray*}
    z^{n+1}\varphi\circ Y^{n+1} &=& z(z^n \varphi\circ Y^n)\circ Y   \\
                                &=& z\bigl(\varphi  \star U_{z\wt R(\varphi)}^n(e)\bigr)\circ Y \\
                                &=& z(\varphi \circ Y) \star  U_{z\wt R(\varphi)}^n(e)
                                        +z \varphi  \star \bigl(U_{z\wt R(\varphi)}^n(e) \circ Y\bigr) \\
                                &=& \varphi  \star \bigl(z\wt R(\varphi) \star U_{z\wt R(\varphi)}^n(e)
                                                                +zU_{z\wt R(\varphi)}^n(e)\circ Y\bigr) \\
                                &=&\varphi \star U_{z\wt R(\varphi)}^{n+1}(e).
\end{eqnarray*}}
\end{proof}

Let us go back to the proof of Proposition~\ref{ck2}. According to
Lemma~\ref{fix} we have for any $t$, at least formally:
\begin{equation}\label{formel}
    \varphi^t = \varphi  \star  \exp(tU_{z\wt R(\varphi)})(e).
\end{equation}
We still have to fix the convergence of the exponential just above
in the case when $z\wt R(\varphi)$ belongs to $\Cal L(\Cal H,\Cal
A_+)$. Let us consider the following decreasing bifiltration of
$\Cal L(\Cal H,\Cal A_+)$~:
$$
    \Cal L_+^{p,q} = \left(z^q\Cal L(\Cal H,\Cal A_+)\right)\cap \Cal L^p,
$$
where $\Cal L^p$ is the set of those $\alpha\in\Cal L(\Cal H,\Cal
A)$ such that $\alpha(x)=0$ for any $x\in\Cal H$ of degree $\le
p-1$. In particular $\Cal L^1=\g g_0$. Considering the associated
filtration~:
$$
    \Cal L_+^n=\sum_{p+q=n}\Cal L_+^{p,q},
$$
we see that for any $\gamma\in\g g_0\cap\Cal L(\Cal H,\Cal A_+)$
the transformation $U_\gamma$ increases the filtration by $1$,
i.e~:
$$
    U_\gamma(\Cal L^n_+)\subset\Cal L^{n+1}_+.
$$
The algebra $\Cal L(\Cal H,\Cal A_+)$ is not complete with respect
to the topology induced by this filtration, but the completion is
$\Cal L(\Cal H,\widehat{\Cal A_+})$, where $\widehat{\Cal
A_+}=\CC[[z]]$ stands for the  formal series. Hence the right-hand
side of (\ref{formel}) is convergent in $\Cal L(\Cal
H,\widehat{\Cal A_+})$ with respect to this topology. Hence for
any $\gamma \in \Cal L(\Cal H,\Cal A_+)$ and for $\varphi$ such
that $z\wt R(\varphi)=\gamma$ we have $\varphi^t=\varphi \star
h_t$ with $h_t\in\Cal L(\Cal H,\widehat{\Cal A_+})$ for any $t$.
On the other hand we already know that $h_t$ takes values in
meromorphic functions for each $t$. So $h_t$ belongs to $\Cal
L(\Cal H,\Cal A_+)$, which proves the first part of the second
assertion. Equation (\ref{gen2}) at $t=0$ reads:
\begin{equation}\label{gen3}
    z\wt R(\varphi)=\dot h(0)=\frac{d}{dt}\restr{t=0}(\varphi^t)_+.
\end{equation}
For $\varphi\in G_{\Cal A_{-}}^{\Phi}$ we have, thanks to the
property $\varphi(\mop{Ker}\epsilon)\subset \Cal A_ -$:
 \allowdisplaybreaks{
\begin{eqnarray*}
    h_t(x)=(\varphi^t)_+(x) &=& (I-\pi)\Big(\varphi^t(x)+\sum_{(x)}\varphi^{{\star}-1}(x')\varphi^t(x'')\Big) \\
                            &=& t(I-\pi)\big( z|x|\varphi(x)+z\sum_{(x)}\varphi^{{\star}-1}(x')\varphi(x'')|x''|\Big)+O(t^2)\cr
                            &=&t\mop{Res}(\varphi\circ Y)+O(t^2),
\end{eqnarray*}}
hence:
\begin{equation}\label{beta2}
    \dot h(0)=\mop{Res }(\varphi\circ Y).
\end{equation}
From equations (\ref{gen}), (\ref{E}) and (\ref{beta2}) we get:
\begin{equation}\label{beta4}
    z\wt R(\varphi)=\mop{Res }(\varphi\circ Y)
\end{equation}
for any $\varphi\in G_{\Cal A_{-}}^\Phi$, hence $z\wt
R(\varphi)\in\g g^c$. Conversely let $\beta$ in $\g g^c$. Consider
$\psi=\wt R^{-1}(z^{-1}\beta)$. This element of $G_{\Cal A}$
verifies by definition, thanks to equation (\ref{E})~:
$$
    z\psi\circ Y=\psi \star \beta.
$$
Hence for any $x\in\mop{Ker}\epsilon$ we have~:
$$
    z\psi(x)={1\over |x|}\Bigl(\beta(x)+\sum_{(x)}\psi(x')\beta(x'')\Bigr).
$$
As $\beta(x)$ is a constant (as a function of the complex variable
$z$) it is easily seen by induction on $|x|$ that the right-hand
side evaluated at $z$ has a limit when $z$ tends to infinity. Thus
$\psi(x) \in \Cal A_-$, and then~:
$$
    \psi =\wt R^{-1}\Big({1\over z}\beta\Big)\in G_{\Cal A_{-}}^\Phi,
$$
which proves assertion (\ref{it:two}). \medskip

Let us prove assertion (\ref{it:three}): Equation
$\varphi^t=\varphi \star  h_t$ together with
$(\varphi^t)^s=\varphi^{t+s}$ yields:
\begin{equation}
    h_{s+t}=h_s \star (h_t)^s.
\end{equation}
Taking values at $z=0$ immediately yields the one-parameter group
property:
\begin{equation}
    F_{s+t}=F_s \star F_t
\end{equation}
thanks to the fact that the evaluation at $z=0$ is an algebra
morphism.
\end{proof}

We can now give a definition of the beta-function~:

\begin{defn}
For any $\varphi\in G_{\Cal A}^\Phi$, the beta-function of
$\varphi$ is the generator of the one-parameter group $F_t$
defined by equation (\ref{rg}) in Proposition \ref {ck2}. It is
the element of the dual $\Cal H^\star $ defined by:
$$
    \beta(\varphi) = \frac{d}{dt}\restr{t=0}F_t(x)
$$
for any $x \in \Cal H$.
\end{defn}

\begin{prop}\label{altbeta}
For any $\varphi\in G_{\Cal A}^\Phi$ the beta-function of
$\varphi$ coincides with the one of the negative part
$\varphi_-^{{\star}-1}$ in the Birkhoff decomposition. It is given
by any of the three expressions:
 \allowdisplaybreaks{
\begin{eqnarray*}
    \beta(\varphi) &=& \mop{Res }\wt R(\varphi) \\
                   &=& \mop{Res }(\varphi_-^{{\star}-1}\circ Y)\\
                   &=& -\mop{Res }(\varphi_-\circ Y).
\end{eqnarray*}}
\end{prop}

\begin{proof}
The third equality will be derived from the second by taking
residues on both sides of the equation:
$$
    0=\wt R(e)=\wt R(\varphi_-)
                     + \varphi_-^{{\star}-1} \star \wt R(\varphi_-^{{\star}-1}) \star \varphi_-,
$$
which is a special instance of the cocycle formula
(\ref{cocycle}). Suppose first $\varphi\in G_{\Cal A_{-}}^{\Phi}$,
hence $\varphi_-^{{\star}-1}=\varphi$. Then $z\wt R(\varphi)$ is a
constant according to assertion~\ref{it:two} of
Proposition~\ref{ck2}. The proposition then follows from equation
(\ref{beta2}) evaluated at $z=0$, and equation (\ref{beta4}).
Suppose now $\varphi\in G_{\Cal A}^\Phi$, and consider its
Birkhoff decomposition. As both components belong to $G_{\Cal
A}^\Phi$ we apply Proposition \ref{ck2} to them. In particular we
have:
 \allowdisplaybreaks{
\begin{eqnarray*}
                    \varphi^t &=& \varphi \star  h_t,\\
    (\varphi_-^{{\star}-1})^t &=& \varphi_-^{{\star}-1}{\star}v_t,\\
                (\varphi_+)^t &=& \varphi_+ \star w_t,
\end{eqnarray*}}
and equality $\varphi^t=(\varphi_-^{{\star}-1})^t \star
(\varphi_+)^t$ yields:
\begin{equation}\label{twist}
    h_t=(\varphi_+)^{{\star}-1} \star v_t \star \varphi_+ \star w_t.
\end{equation}
We denote by $F_t, V_t, W_t$ the one-parameter groups obtained
from $h_t,v_t,w_t$, respectively, by letting the complex variable
$z$ go to zero. It is clear that $\varphi^+\restr{z=0}=e$, and
similarly that $W_t$ is the constant one-parameter group reduced
to the co-unit $\varepsilon$. Hence equation (\ref{twist}) at
$z=0$ reduces to:
\begin{equation}\label{Ft=Vt}
    F_t=V_t,
\end{equation}
hence the first assertion. the cocycle equation (\ref{cocycle})
applied to the Birkhoff decomposition reads:
$$
    \wt R(\varphi)=\wt R(\varphi_+)+(\varphi_+)^{{\star}-1} \star \wt R(\varphi_-^{{\star}-1}) \star \varphi_+.
$$
Taking residues of both sides yields:
$$
    \mop{Res }\wt R(\varphi)=\mop{Res } \wt R(\varphi_-^{{\star}-1}),
$$
which ends the proof.
\end{proof}

\begin{defn} \label{def:Ft=Vt}
The one-parameter group $F_t=V_t$ above is the
{\sl{renormalization group}\/} of $\varphi$ \cite{CK3}.
\end{defn}

\begin{rmk} {\rm{
It is possible to reconstruct $\varphi_-$ from $\beta(\varphi)$
using a scattering-type formula (\cite{CK3,Co,CM3,Ma01}). Hence
$\varphi_-$ (i.e. the divergence structure of $\varphi$) is
uniquely determined by its residue.}}
\end{rmk}


\subsection{The matrix representation of the grading derivation}
\label{ssect:matrixY}

As in reference \cite{CK3}, denote by $Z_0$ the derivation of the
algebra $\Cal L(\Cal H,\Cal A)$ (which is also a derivation of the
the Lie algebra $\g g_{\Cal A}$) given by $\alpha \mapsto \alpha
\circ Y$. Let $\wt{\Cal L}$ be the semi-direct product of $\Cal
L(\Cal H,\Cal A)$ with $Z_0$, and let $\wt{\g g_{\Cal A}}$ be the
semi-direct product $\g g_{\Cal A} \semi\CC.Z_0$. Similarly let
$\wt G_{\Cal A}=G_{\Cal A}\semi\CC$ be the semi-direct product of
the group $G_{\Cal A}$ by the one-parameter group $\theta_t=\exp
tZ_0$ of automorphisms. Let $J$ be any {\sl graded\/} left coideal
of $\Cal H$. We suppose further that the filtration-ordered basis
$(x_i)_{i\in I}$ of $J$ is graded. i.e. made of homogeneous
elements. The degree of $x_i$ will be denoted by $|x_i|$. We want
to extend the matrix representation $\Psi_J$ from $\Cal L(\Cal
H,\Cal A)$ to $\wt{\Cal L}$. This can be done by representing
$Z_0$ by a diagonal matrix.

\begin{prop}\label{extension}
The correspondence $\Psi_J: \wt{\Cal L} \to \mop{End}_{\Cal
A}(\Cal A\otimes J)$ defined as in Paragraph \ref{ssect:matrix} on
$\Cal L(\Cal H,\Cal A)$, and such that:
\begin{equation}\label{zzero}
    \Psi_J[Z_0](x_i)=|x_i|.x_i
\end{equation}
is an algebra morphism.
\end{prop}

\begin{proof}
We only have to show the equality\footnote{There is a minus sign
in front of $Z_0$ due to the fact that we have put it on the right
inside of the bracket, reflecting the fact that the action of the
one-parameter group has been written on the right.}:
\begin{equation}\label{morph}
    \big[\Psi_J[f],\, \Psi_J[-Z_0]\big] = \Psi_J([f,-Z_0]) =\Psi_J(f\circ Y).
\end{equation}
This follows by a direct computation:
 \allowdisplaybreaks{
\begin{eqnarray*}
    \big[\Psi_J[f],\,
    \Psi_J[-Z_0]\big](x_j) &=& -\Psi_J[f]\Psi_J[Z_0](x_j)+\Psi_J[Z_0]\Psi_J[f](x_j)\\
                           &=& \sum_i(|x_i|-|x_j|)f(M_{ij})\otimes x_i,
\end{eqnarray*}}
whereas:
 \allowdisplaybreaks{
\begin{eqnarray*}
 \Psi_J[f\circ Y](x_j) &=& \sum_i(f\circ Y)(M_{ij})\otimes x_i\\
                       &=& \sum_i|M_{ij}|f(M_{ij})\otimes x_i.
\end{eqnarray*}}
By definition of the coproduct matrix, and thanks to the fact that $\Cal H$ is
graded, the coefficients $M_{ij}$ are homogeneous of degree $|x_i|-|x_j|$, which
finishes the proof.
\end{proof}

By taking exponentials of the above diagonal matrices, we of
course get a matrix representation of the one-parameter group
$\theta_{t}$ (\ref{group}), namely:
\begin{equation}\label{expzzero}
    \Psi_J[\exp tZ_0](x_i)=e^{t|x_i|}x_i.
\end{equation}


\subsection{Matrix differential equations}
\label{ssect:matrixDiffEqs}

We fix a graded coideal $J$ of $\Cal H$, and we therefore
introduce the following notation:
\begin{equation}\label{notationpsi}
    \wh f:=\Psi_J[f].
\end{equation}
Keeping the notations of Paragraph \ref{subsect:beta} we have
then, as a consequence of Propositions~\ref{morphism} and
\ref{extension}:
\begin{equation}\label{action-group}
    \wh{\varphi^t}(z) = e^{tz\wh{Z_0}} \wh\varphi(z) e^{-tz\wh{Z_0}},
\end{equation}
as an equality of size $|I|$ square matrices with coefficients in
meromorphic functions of the complex variable $z$. With respect to
the example left coideal generated by $\Cal T' \subset \Cal T$ in
(\ref {6space}) we find:
\begin{equation}
\label{Z0-Matrix}
    \wh{Z_0}=  {\small{
\left(
  \begin{array}{cccccc}
   0           & 0           & 0           & 0         &0 \\
   0           & 1           & 0           & 0         &0 \\
   0           & 0           & 2           & 0         &0 \\
   0           & 0           & 0           & 3         &0 \\
   0           & 0           & 0           & 0         &4
  \end{array}
           \right)}}
\;\; \makebox{ and }\;
 e^{tz\wh{Z_0}} =
  {\small{
\left(
  \begin{array}{cccccc}
   1_{\Cal A}  & 0           & 0           & 0         &0 \\
   0           & e^{zt}      & 0           & 0         &0 \\
   0           & 0           & e^{2zt}     & 0         &0 \\
   0           & 0           & 0           & e^{3zt}   &0 \\
   0           & 0           & 0           & 0         & e^{4zt}
  \end{array}
           \right)}}
\end{equation}

\noindent We now change the general notations slightly and
consider $\wh\varphi$ as a function of the variable $t \in \CC$
with values in $\Cal A \otimes J$ (i.e. as a matrix-valued
function of both variables $(t,z)$). More precisely we put:
$$
    \wh\varphi(t,z)=e^{tz\wh{Z_0}}\wh\varphi(0,z)e^{-tz\wh{Z_0}},
$$
where $\wh\varphi(0,z)$ stands for the old $\wh\varphi(z)$. Now
$\wh\varphi_-(t)$ (resp. $\wh\varphi_+(t)$) will stand for the
negative (resp. positive) component of the Birkhoff decomposition
of $\wh\varphi(t)$  (\ref{matrixBirkhoff}), for any $t\in\CC$.
Using the toy-model character $\varphi=\varphi(s,z)$ in
equation~(\ref{toycharacter}) with parameterized 't~Hooft mass, we
find explicitly:
 \allowdisplaybreaks{
\begin{eqnarray}
 \wh{\varphi^t}(z) = \wh{\varphi}(t,z) &=& {\small{\left(
  \begin{array}{cccccc}
   1_{\Cal A}                       & 0                            & 0                              & 0                          &0 \\
    (e^{-t}\alpha(0))^{-z}B_1       & 1_{\Cal A}                   & 0                              & 0                          &0 \\
    (e^{-t}\alpha(0))^{-2z}B_2B_1   & (e^{-t}\alpha(0))^{-z}B_1    & 1_{\Cal A}                     & 0                          &0 \\
    (e^{-t}\alpha(0))^{-3z}B_3B^2_1 & (e^{-t}\alpha(0))^{-2z}B^2_1 & 2 (e^{-t}\alpha(0))^{-z}B_1    & 1_{\Cal A}                 &0 \\
    (e^{-t}\alpha(0))^{-4z}B_4B^3_1 & (e^{-t}\alpha(0))^{-3z}B^3_1 & 3 (e^{-t}\alpha(0))^{-2z}B^2_1 & 3(e^{-t}\alpha(0))^{-z}B_1 & 1_{\Cal A}
  \end{array}
           \right)}} \nonumber\\
     &=&
   e^{tz\wh{Z_0}}
 {\small{\left(
  \begin{array}{cccccc}
   1_{\Cal A}      & 0            & 0              & 0            &0 \\
   \alpha(0)^{-z}B_1       & 1_{\Cal A}   & 0              & 0            &0 \\
   \alpha(0)^{-2z}B_2B_1   & \alpha(0)^{-z}B_1    & 1_{\Cal A}     & 0            &0 \\
   \alpha(0)^{-3z}B_3B^2_1 & \alpha(0)^{-2z}B^2_1 & 2 \alpha(0)^{-z}B_1    & 1_{\Cal A}   &0 \\
   \alpha(0)^{-4z}B_4B^3_1 & \alpha(0)^{-3z}B^3_1 & 3 \alpha(0)^{-2z}B^2_1 & 3\alpha(0)^{-z}B_1   & 1_{\Cal A}
  \end{array}
           \right)}}
    e^{-tz\wh{Z_0}} \label{matrixeq}
\end{eqnarray}}
Hence, we observe the simple transformation on
$\wh{\varphi}(z)=\wh{\varphi}(t,z)$, where $t$ parameterizes
't~Hooft's unit mass (\ref{parameter}), i.e.,
$\alpha(t):=\frac{a}{e^t\mu}=e^{-t}\alpha(0)>0$:
$$
    \wh{\varphi}^0(z):=\wh{\varphi}(0,z) \xrightarrow{Ad[e^{tz\wh{Z_0}}]} \wh{\varphi}^t(z)=\wh{\varphi}(t,z).
$$
We introduce the auxiliary matrix:
\begin{equation}\label{aux}
    A:=\wh\varphi e^{tz\wh{Z_0}},
\end{equation}
as well as its Birkhoff decomposition:
\begin{equation}\label{birkhoffA}
    A=A_-^{-1} A_+,\ \hbox{ \rm with }\ A_-=\wh\varphi_-
                                      \hbox{ \rm and }
                                        A_+=\wh\varphi_+ e^{tz\wh{Z_0}}.
\end{equation}
From the obvious equality:
$$
    A(t) = e^{tz\wh{Z_0}} A e^{-tz\wh{Z_0}}
         = e^{tz\wh{Z_0}}A(0)
$$
we get by differentiating with respect to $t$:
\begin{equation}\label{equadiff1}
     {d \over dt}A = \dot A = z \wh {Z_0} A.
\end{equation}
Using the Birkhoff decomposition of $A$ then yields:
\begin{equation}\label{equadiff2}
    z\wh{Z_0}\wh{\varphi}_-^{\ -1} A_+ = \dot{\wh{\varphi}_-^{\ -1}} A_+ + \wh{\varphi}_-^{\ -1}\dot{A_+}.
\end{equation}
Multiplying both sides with $\wh{\varphi}_-$ on the left and with
$A_+^{-1}$ on the right we get:
\begin{equation}\label{equadiff3}
    \wh{\varphi}_- (z\wh{Z_0}) \wh{\varphi}_-^{\ -1}
                          =\wh{\varphi}_- \dot{\wh{\varphi}_-^{\ -1}} + \dot {A_+} A_+^{-1}.
\end{equation}
Suppose now that $\varphi(0)$ belongs to $G_{\Cal A}^\Phi$, which
implies for the corresponding matrix $\wh
\varphi(0)=\wh\varphi_-^{-1}\wh\varphi_+$:
\begin{equation}\label{gphi}
    {d \over dt}\wh \varphi_- = \dot{\wh \varphi_-}=0.
\end{equation}
Then equation (\ref{equadiff3}) reduces to:
\begin{equation}\label{equadiff4}
    \wh{\varphi}_- (z\wh{Z_0}) \wh{\varphi}_-^{\ -1} = \dot {A_+} A_+^{-1}.
\end{equation}


\subsection{The renormalization group and the beta-function in the matrix setting}
\label{ssect:beta-Matrix}

Keeping the notations of the previous paragraph, it is clear that
if $\varphi(0)$ belongs to $G_{\Cal A}^\Phi$, then $\varphi(t)\in
G_{\Cal A}^\Phi$ for any $t$, and moreover the renormalization
group and the beta-function of $\varphi(t)$ do not depend on $t$.
We can then talk about the renormalization group and the
beta-function of $\varphi$ without mentioning a particular value
of $t$.

\begin{thm}\label{matrixbeta}
The matrix representation of the beta-function reads:
\begin{equation}
 \label{betaMatrix}
    \wh\beta(\varphi) = \wh\varphi_- (z\wh{Z_0}) \wh\varphi_-^{-1} - z \wh{Z_0}.
\end{equation}
\end{thm}

\begin{proof}
This is a direct computation of (scalar-valued) matrices, see
Eq.~(\ref{Ft=Vt}) and Definition~\ref{def:Ft=Vt}:
\begin{equation}
 \wh F_t = \mopl{lim}_{z\to 0}\Big(\wh\varphi_-(z) e^{zt\wh{Z_0}}\wh\varphi_-^{-1}(z) e^{-zt\wh{Z_0}}\Big).
\end{equation}
The limit exists by Proposition~\ref{ck2}. The term inside the
bracket on the right-hand side is holomorphic at zero as a
function of $z$, and so is its derivative with respect to $t$. The
operation $\frac{\partial}{\partial t}$ commutes then with
evaluating at $z=0$, and we get by definition of the
beta-function:
 \allowdisplaybreaks{
\begin{eqnarray*}
    \wh\beta(\varphi) &=& \mopl{lim}_{z\to 0}\frac{\partial}{\partial t}
                           \Big(\wh\varphi_-(z) e^{zt\wh{Z_0}} \wh\varphi_-^{-1}(z) e^{-zt\wh{Z_0}}\Big)\restr{t=0}\\
                      &=& \mopl{lim}_{z\to 0}\Big(\wh\varphi_-(z) z\wh{Z_0} \wh\varphi_-^{-1}(z) - z\wh{Z_0}\Big).
\end{eqnarray*}}
Now, subtracting $z \wh{Z_0}$ on both sides of Equation
(\ref{equadiff4}) gives an expression on the left-hand side which
admits a limit when $z\to\infty$, and a term on the right-hand
side which admits a limit when $z \to 0$. Hence the term:
$$
    \wh\varphi_-(z) z\wh{Z_0}\wh\varphi_-^{-1}(z) -
    z\wh{Z_0}=\wh\varphi_-(z)\big[\wh\varphi_-^{-1}(z),-z\wh{Z_0}\big]
$$
is a matrix with constant coefficients, which proves the theorem.
\end{proof}

Coming back to Remark~\ref{bch-CHI} we find immediately for the
matrix beta-function the simple equation in the Lie algebra
$\wh{\g g}_{\Cal A}$, in accordance with results in \cite{Ma01}:
 \allowdisplaybreaks{
\begin{eqnarray}
      \wh\beta(\varphi) &=& \exp\big(\mathrm{R}(\chi(\widehat{Z}))\big)
                            (z\wh{Z_0})
                           \exp\big(-\mathrm{R}(\chi(\widehat{Z}))\big)
                            - z \wh{Z_0} \nonumber\\
                        &=&
                        z\big[\mathrm{R}(\chi(\widehat{Z})),\wh{Z_0}\big]
                        + \frac{z}{2!}\big[\mathrm{R}(\chi(\widehat{Z})),[\mathrm{R}(\chi(\widehat{Z})),\wh{Z_0}]\big]
                        + \frac{z}{3!}\big[\mathrm{R}(\chi(\widehat{Z})),[\mathrm{R}(\chi(\widehat{Z})),
                                               [\mathrm{R}(\chi(\widehat{Z})),\wh{Z_0}]]\big] + \cdots \nonumber\\
                        &=& z\sum_{n>0} \frac{1}{n!} \mathrm{ad}\big[\mathrm{R}(\chi(\widehat{Z}))\big]^{(n)}(\wh{Z_0})
                         \label{chibeta}
\end{eqnarray}}

The following corollary is a direct consequence of
Equation~(\ref{equadiff4}).

\begin{cor}\label{AprimeAinv}
\begin{equation}
    \dot {A_+} A_+^{-1} = \wh\beta(\varphi) + z\wh{Z_0}.
\end{equation}
\end{cor}
\noindent and for the renormalization matrix we get

\begin{cor}
\begin{equation}
    \wh\varphi_+(t,z) = e^{t(\wh\beta(\varphi) + z\wh{Z_0})} \wh\varphi_+(z,0) e^{-tz\wh{Z_0}}.
\end{equation}
\end{cor}

\begin{proof}
From Corollary~\ref{AprimeAinv} we get:
$$
    A_+(t)=e^{t(\wh\beta(\varphi) + z\wh{Z_0})}A_+(0),
$$
which immediately proves the claim. Alternatively one readily
observes that:
 \allowdisplaybreaks{
\begin{eqnarray*}
    \wh\varphi_+(t,z)  = \wh\varphi_{-} \wh{\varphi^t}
                      &=& \wh\varphi_{-} e^{tz\wh{Z_0}} \wh\varphi(0,z) e^{-tz\wh{Z_0}}\\
                      &=& \wh\varphi_{-} e^{tz\wh{Z_0}} \wh\varphi_{-}^{-1} \wh\varphi_{+}(0,z) e^{-tz\wh{Z_0}}\\
                      &=& e^{t\wh\varphi_- (z\wh{Z_0}) \wh\varphi_-^{-1} } \wh\varphi_+(z,0) e^{-tz\wh{Z_0}}\\
                      &=& e^{t(\wh\beta(\varphi) + z\wh{Z_0})} \wh\varphi_+(z,0) e^{-tz\wh{Z_0}},
\end{eqnarray*}}
where we used the well-known fact
$\exp(A)\exp(B)\exp(-A)=\exp(\exp(A)B\exp(-A))$.
\end{proof}

\begin{cor}\label{scatt} {\rm (Connes--Kreimer's scattering-type formula~\cite{CK1})}
\begin{equation}
    \wh\varphi_- = \mopl{lim}_{t \to +\infty} e^{-t(\frac{\wh\beta(\varphi)}{z} + \wh{Z_0})} e^{t\wh{Z_0}}.
\end{equation}
\end{cor}

\begin{proof}
We again adapt Sakakibara's computation (\cite{Sa04} \S\ 2) to our matrix setting:
 \allowdisplaybreaks{
\begin{eqnarray*}
    e^{-t(\frac{\wh\beta(\varphi)}{z}+\wh{Z_0})} e^{t\wh{Z_0}}
                &=& e^{-t(\wh\varphi_- \wh{Z_0} \wh\varphi_-^{-1})} e^{t\wh{Z_0}}\\
                &=& \wh\varphi_- e^{-t\wh{Z_0}} \wh\varphi_-^{-1} e^{t\wh {Z_0}}\\
                &=& \Psi_J\big[\varphi_- \star\theta_{-t}\big(\varphi_-^{{\star}-1}\big)\big].
\end{eqnarray*}}
Now we have for any homogeneous $x \in \Cal H$ of degree $\ge 1$:
 \allowdisplaybreaks{
\begin{eqnarray*}
    \mopl{lim}_{t \to +\infty}\big(\varphi_- \star \theta_{-t}(\varphi_-^{{\star}-1})\big)(x)
       &=& \mopl{lim}_{t \to +\infty}\Big(\varphi_-(x) + e^{-t|x|}\varphi_-^{-1}(x)
                                                  + \sum_{(x)}\varphi_-(x')e^{-t|x''|}\varphi_-^{-1}(x'')\Big)\\
       &=& \varphi_-(x).
\end{eqnarray*}}
Replacing $x$ with the matrix coefficients $M_{ij}$ proves then the corollary.
\end{proof}

\noindent This result becomes evident on the level of matrices
when going back to equation~(\ref{matrixeq}). Assume for a moment
$z \in \mathbb{R}$ positive. One observes by replacing $t$ by $-t$
that in the first equality on the right-hand side each entry has
the form $(\wh {\varphi^{-t}}_{ij})_{i\geq
j}=(\exp(-tz|M_{ij}|)\wh{\varphi}_{ij})_{i\geq j}$ with $M_{ij}$
being the coproduct matrix in (\ref{coprodMat6}). Hence, with
$|M_{ii}|=0$ we see immediately that $(\wh {\varphi^{-t}}_{ij})
\xrightarrow{t \to \infty} {\bf{1}}$.

\begin{rmk}{\rm{
Considering the Proposition~\ref{altbeta} in the matrix setting we
have an alternative matrix representation of the beta-function:
\begin{equation}\label{altbeta2}
    \wh\beta(\varphi)=[\mop{Res }\wh\varphi_-,\,\wh Z_0].
\end{equation}}}
\end{rmk}





\vspace{0.5cm} {\emph{Acknowledgements}}: The first author
acknowledges greatly the support by the European Post-Doctoral
Institute and Institut des Hautes \'Etudes Scientifiques
(I.H.\'E.S.). He profited from discussions with D.~Kreimer. The
second author greatly acknowledges constant support from the
Centre National de la Recherche Scientifique (C.N.R.S.).


\end{document}